\documentclass[apj]{emulateapj}
\usepackage{color}
\usepackage{epsf}
\usepackage{ulem}
\usepackage{graphicx}

\newcommand{\Om}{\Omega_m}

\newcommand{\OL}{\Omega_\Lambda}

\newcommand{\Oh}{\Omega_m h^2}
\newcommand{\Obhh}{\Omega_b h^2}

\newcommand{\hMpc}{h^{-1}{\rm\;Mpc}}

\newcommand{\trihGpc}{h^{-3}{\rm\;Gpc^3}}

\newcommand{\ihMpc}{h{\rm\;Mpc^{-1}}}

\newcommand{\al}{\alpha}
\newcommand{\sial}{\sigma_\alpha}

\newcommand{\FoG}{F_{\rm fog}}
\newcommand{\Pm}{P_m}
\newcommand{\nPt}{nP_{0.2}}
\newcommand{\sig}{\sigma}
\newcommand{\Sigxy}{\Sigma_{\rm xy}}
\newcommand{\Sigz}{\Sigma_{z}}
\newcommand{\Nb}{{\it N}-body}
\newcommand{\Signl}{\Sigma_{\rm m}}
\newcommand{\Signn}{\Sigma_{\rm nl}}
\newcommand{\Pano}{A(k)}
\newcommand{\Plin}{P_{\rm lin}}
\newcommand{\Pres}{P_{\rm res}}
\newcommand{\Pnl}{P_{\rm nl}}
\newcommand{\Pobs}{P_{\rm obs}}
\newcommand{\Zel}{Zel'dovich}
\newcommand{\rechi}{\chi^2/{\rm DOF}}
\newcommand{\don}{d_1}
\newcommand{\dt}{d_2}

\begin{document}
\title{Non-linear structure formation and the acoustic scale}

\author{
Hee-Jong Seo\altaffilmark{1,2},
Ethan R.\ Siegel\altaffilmark{2},
Daniel J.\ Eisenstein\altaffilmark{2},  
Martin White\altaffilmark{3,4}}
\submitted{Accepted for publication in the Astrophysical Journal}

\begin{abstract}
We present high signal-to-noise measurements of the acoustic scale in
the presence of nonlinear growth and redshift distortions using $320
\trihGpc$ of cosmological PM simulations. Using simple fitting methods,
we obtain robust measurements of the acoustic scale with scatter close
to that predicted by the Fisher matrix.  We detect and quantify the shift in the
acoustic scale by analyzing the power spectrum: we detect at greater than
$5\sig$ a decrease in the acoustic scale in the real-space matter power
spectrum of 0.2\% at $z=1.5$, growing to 0.45\% at $z=0.3$. In redshift
space, the shifts are about 25\% larger: we detect a decrease of 0.25\%
of at $z=1.5$ and 0.54\% at $z=0.3$. Despite the nonzero amounts, these
shifts are highly predictable numerically, and hence removable within
the standard ruler analysis of clustering data. Moreover, we show that
a simple density-field reconstruction method substantially reduces the
scatter and nonlinear shifts of the acoustic scale measurements: the
shifts are reduced to less than 0.1\% at $z=0.3-1.5$, even in the presence
of non-negligible shot noise. Finally, we show that the ratio of the 
cosmological distance to the sound horizon that would be inferred from 
these fits is robust
to variations in the parameterization of the fitting method and reasonable
differences in the template cosmology.
\end{abstract}


\altaffiltext{1}{Center for Particle Astrophysics, Fermi National Accelerator Laboratory, P.O. Box 5
00, Batavia, IL 60510-5011; sheejong@fnal.gov}
\altaffiltext{2}{Steward Observatory, University of Arizona,
                933 N. Cherry Ave., Tucson, AZ 85121}
\altaffiltext{3}{Departments of Physics and Astronomy, University of California, Berkeley, CA94720}
\altaffiltext{4}{Lawrence Berkeley National Laboratory, 1 Cyclotron Road, Berkeley, CA}

\section{Introduction}
Baryon acoustic oscillations (BAO) imprint a distinctive feature in the large-scale structure of both the cosmic microwave background and the matter density 
field. The sound waves propagating through the primeval plasma of photons 
and baryons in the early Universe freeze out 
when the photons are freed from baryons at the epoch of
recombination. This leaves unique oscillatory features in the CMB
\citep[e.g.,][]{Mil99,deB00,Han00,Lee01,HalDasi,Netter02,Pearson03,BenoitArcheops,BennettWmap,Hinshaw07,Hinshaw08}
as well as a lower contrast
counterpart in the large-scale structure of matter
\citep[e.g.,][]{Peebles70,SZ70,Bond84,Holtzman89,HS96,Hu96,EH98,Meiksin99,Eisen05,Cole05,Hutsi06,Tegmark06,Percival07a,Percival07b,Blake07,Pad07,Okumura08,Estra08}.
The physical scale of the BAO is set by the distance that the sound
waves have traveled before recombination, known as the sound horizon or
acoustic scale. Since the sound horizon scale can be well determined by
present and future CMB data, the BAO can serve as an excellent standard
ruler \citep{Hu96,EHT98,Eisen03,Blake03,Hu03,SE03}. Observing the BAO
from large-scale structure at different redshifts with the knowledge
of their true physical scale allows us to geometrically measure the
expansion history of the Universe and thereby identify the driving
force behind the observed accelerated expansion, i.e., dark energy
\citep[e.g.,][]{Riess98,Perlm99}.

Using the BAO as a standard ruler for precision cosmology requires that
we understand all of the physical effects that could alter the acoustic
scale during the evolution of the Universe. There are two important
aspects of performance to consider for a standard ruler measurement
with BAO: precision and accuracy.  The precision of the standard ruler
method crucially depends on the nonlinear evolution of structure growth
and observational effects, such as redshift distortions and galaxy
bias. With time, these nonlinear effects increasingly degrade the contrast
of the BAO in the matter power spectrum and correlation function,
decreasing the signal-to-noise of the standard ruler method. Recently,
significant progress has been made on modeling the nonlinear evolution
of the amplitude of the BAO with analytic and/or numerical methods
\citep{Meiksin99,Springel05,Angulo05,SE05,Angulo05,White05,Jeong06,Crocce06b,ESW07,Huff07,Smith07,Matarrese07,Smith08,Angulo08,Crocce08,Mat08,Taka08,Sanchez08}
and on forecasting the resulting
signal-to-noise for future galaxy redshift surveys
\citep[][]{Blake03,Linder03,Hu03,SE03,Cooray04,Mat04,Amen05,Blake05,Glazebrook05,Dolney06,Zhan06,Blake06,SE07}.
We can now make reasonable predictions for the degradation of the BAO
due to nonlinear growth and redshift distortions.

The accuracy of the standard ruler method depends on the calibration
of the physical scale of the BAO, i.e., the sound horizon. Future BAO
surveys aim to push the standard ruler method to near the cosmic variance
limit. This will require a calibration of the sound horizon scale to
about 0.1\% accuracy. However, nonlinear evolution not only degrades
the precision of the standard ruler method by decreasing the contrast
of the BAO, but also may alter [i.e., shift] the observed BAO scale in
the large-scale structure at low redshift. Such a shift, relative to the
linear acoustic scale derived from the CMB, will degrade the accuracy of
the ruler. Failure to appropriately account for such a shift will result
in biased measurements of the cosmological distances and therefore bias
in the inferred dark energy parameters.

Previously, \citet{SE05} had no detection of a shift (or
bias) in the acoustic scale at the $\sim 1\%$ level in the
nonlinear real-space power spectrum from N-body data 
\citep[also see][]{Huff07,Ma07,ESW07,SE07,Angulo08,Sanchez08}. Meanwhile,
\citet{Smith08} predicts more than a percent level shift in the acoustic
scale, when defining the observed scale as the local maxima of the peak
in the correlation function \citep[also see][]{Smith07,Guzik07}. However,
the effect of the broadband shape, as well as the effect of the nonlinear
smoothing of the BAO, on the shift of the acoustic scale can be largely
marginalized over with appropriate parameterizations, which is probably
the reason why such a large shift is not detected in \citet{SE05} and
\citet{Angulo08}. That is, the choice of acoustic scale estimator matters;
one cannot quantify a nonlinear shift without defining the estimator and
one would surely prefer to use one that is close to optimal. The latter
goal recommends template fitting to a significant portion of the power
spectrum or correlation function, rather than peak finding methods.

Recent analytic and numerical work by \citet{Crocce08} makes use of
renormalization perturbation theory and predicts a nonzero, irreducible
shift at the few tenths of a percent level in real space due to 
nonlinear mode coupling. This is referred to as a ``mode-coupling
shift''; a similar context is introduced as ``physical shifts'' in
\citet{Smith08}. The exact level of this mode-coupling shift would also
depend on estimators: for example, the effects of the mode-coupling
shifts that are different on different nodes may average out to some
degree in the case of a global fit over several acoustic peaks. The good
estimators will be ones that marginalize over most of the effects from
the broadband shape and minimize contributions from the mode-coupling
shifts. Any residual shifts can be modeled through numerical computation,
such as a study of a large volume of \Nb\ data.

In this paper, we show that most of the acoustic scale information can
be extracted via a template fitting method. In order to test and confirm such
a low level mode-coupling shift in redshift space as well as in real
space, we require a very large volume of cosmological simulation. We use
$320\trihGpc$ of simulations to test the precision and the sub-percent
level accuracy of the standard ruler method with BAO.  In this paper,
we consider both nonlinear growth and redshift distortions. The effect
of galaxy bias will be studied in our subsequent paper (Siegel et al.,
in preparation).

\citet{ESSS07} show that a portion of the nonlinear degradation of the
BAO can be undone with a simple reconstruction scheme based on the \Zel\
approximation \citep[also see][]{SE07}. We test how well we can undo
the nonlinear shift as well as the erasure of the BAO, even given the
presence of shot noise.

In \S~\ref{sec:method}, we describe our cosmological N-body simulations
and methods of $\chi^2$ analysis to measure the acoustic scales from the
simulations. In \S~\ref{sec:nlgrowth}, we present the resulting shifts and
errors on the measurements of the acoustic scale when accounting for the
nonlinear growth. In \S~\ref{sec:reddis}, we proceed to redshift space,
and present the effects of redshift distortions. In \S~\ref{sec:recons},
we show that not only the nonlinear erasure but also the nonlinear shift
on the acoustic scale can largely be undone via reconstruction. In
\S~\ref{sec:fisher}, we compare our results of signal-to-noise with
the Fisher matrix calculations.  In \S~\ref{sec:robust}, we test how
our results depend on the small errors on the template cosmology used
in our fitting. Finally, in \S~\ref{sec:disc}, we summarize the major
results obtained in this paper and point towards future directions for
probing dark energy using BAO as a standard ruler.

\section{Simulations and the methods of analysis}\label{sec:method}
We have chosen to investigate a specific cosmology of the $\Lambda$CDM
family: $\Om=0.25$, $\OL=0.75$, $\Obhh=0.0224$, $h=0.7$, $n=0.97$ and
$\sig_8=0.8$.  In order to model the non-linear evolution of structure,
we make use of N-body simulations.  Since our focus is on large scales
and we need a lot of volume, we have elected to run 40 realizations
of a $2\,h^{-1}$Gpc box using a parallel particle-mesh (PM) code.
Each simulation employs $1024^3$ particles on a $2048^3$ force mesh,
for a mesh resolution of $1\,h^{-1}$Mpc and a particle mass of $5\times
10^{11}\,h^{-1}M_\odot$.  The acoustic scale is thus resolved by $\sim
100$ grid cells in each dimension and the non-linear scale by $\sim
10$ cells.  The initial conditions were generated using the Zel'dovich
approximation \citep{Zel70} starting from a regular grid at $z=50$.
The linear theory power spectrum for the initial conditions was computed
by evolution of the coupled Einstein, fluid and Boltzmann equations using
the code described in \citet{Whi96}.  \citet{SSWZ} find that this code
agrees well with CMBfast \citep{SeZa96}.  We used constant steps in $\ln
a$ with steps of either $5\%$ or $10\%$.  The larger time steps resulted
in a loss of power at small scales, but we saw no dependence on time step
in the acoustic scale information.

Power spectra were computed at $z=1.5$, 1.0, 0.7, and 0.3 by assigning the
mass to a grid using the cloud-in-cell interpolation and using wavenumber
bins of width $\Delta k=0.0047\,h\,{\rm Mpc}^{-1}$. In order to reduce
the data volume that we need to store and manipulate, we save only 1\%
of the processed data for the later use (\S~\ref{sec:recons}). In this
paper, we use this 1\% fraction only for our studies of reconstruction
and shot noise.

We perform a $\chi^2$ analysis to fit the spherically averaged power spectrum 
$\Pobs(k)$ to template power spectra $\Pm(k/\al)$. We parameterize the 
observed power spectra as
\begin{equation}\label{eq:Pobs}
\Pobs(k)=B(k)\Pm(k/\al)+\Pano,
\end{equation}
where $\al$, $B(k)$, and $\Pano$ are fitting parameters. Here $\al$
is a scale dilation parameter and represents the ratio of the true
(or linear) acoustic scale to the measured scale. For example, $\al>1$
means that the measured BAO being shifted toward larger $k$ relative to
the linear power spectrum. In the case we fail to account for any
nonlinear shift on the acoustic scale in the standard ruler method,
$\al$ represents the ratio of the mis-measured distance to the true
distance. The term $B(k)$ allows a scale-dependent nonlinear growth,
and $\Pano$ represents an anomalous power, i.e., additive terms from the
nonlinear growth and shot noise. By including both $B(k)$ and $\Pano$
with a large number of parameters, we minimize the contribution to the
standard ruler method from the broadband shape of the power spectrum. We
try various parameterizations for $B(k)$ and $\Pano$, as described in
\S~\ref{sec:nlgrowth} and \S~\ref{sec:reddis}.

We calculate the model or template power spectrum $\Pm$ by
modifying the BAO portion of the linear power spectrum with
a nonlinear parameter $\Signl$ to account for the degradation
of the BAO due to nonlinear effects and redshift distortions
\citep{Eisen05,Tegmark06,Crocce06b,ESSS07,Crocce08,Mat08}:
\begin{eqnarray}\label{eq:Pmodel}
\Pm(k)&=&\left[ \Plin(k)-P_{\rm smooth}(k)\right] \exp{\left[ -k^2 \Signl^2/2 \right]} \nonumber \\
&&+P_{\rm smooth}(k),
\end{eqnarray}
where $\Plin$ is the linear power spectrum and $P_{\rm smooth}$ is the nowiggle form from \citet{EH98}. Any details
of physics overlooked in the template power spectrum will be, at least
partly, absorbed into $B(k)$ and $\Pano$. In fact, we will show that
our results do not vary for a wide range of $\Signl$, as the fitting
formula can alter the amplitude of the BAO by trading power between
$B(k)P_m$ and $\Pano$, as $B(k)$ and $\Pano$ have enough flexibility.
Fine-tuning the choice of $\Signl$ in $\Pm$ does not affect our results, and
therefore we do not include $\Signl$ as a fitting parameter. We use a
fitting range of $0.02\ihMpc \leq k \leq 0.35\ihMpc$.

As the true covariance matrix of the \Nb\ simulations is unknown a priori,
we use the variation between the simulations to assess the true scatter
in $\al$. We fit each simulation assuming the covariance matrix for $P(k)$ 
is that
of a Gaussian random field, i.e., assuming independent band powers with
variances determined by the number of independent modes in each. While
this is not an optimal weighting of the data for the determination of
$\al$, the effects from non-Gaussianity in the density field will still 
be reflected in the scatter between best-fit $\al$'s from different simulations.

We use different groupings to assess the scatter between our
simulations. Our primary approach is to measure the scatter in $\al$
from jackknife sampling of the 40 simulations. However, we have also
used bootstrap sampling for 40 sets of 1 simulation. The results from
the two approaches are highly consistent; we will quote results for the
jackknife case. Note that the scatter in $\al$, i.e., $\sial$, that we
quote represents the scatter associated with the mean value of $\al$
for a total volume of $320\trihGpc$, not the scatter between samples.


\section{A real-space analysis of the BAO}\label{sec:nlgrowth}
\subsection{Non-linear Growth of Structure}

Nonlinear growth arises when the density perturbation on a given scale
reaches an amplitude of order unity; the evolution of perturbations
of different wave modes becomes increasingly coupled with one
another, causing a substantial departure from linear evolution
\citep[e.g.,][]{Jusz81,Vish83,Makino92,Jain94,Bhara96a,Bhara96b,MW99,Sco99}. Such
a nonlinear process erases the BAO of the power spectrum
\citep[e.g.,][]{Meiksin99,SE05}, thereby degrading the signal of the
standard ruler method. It also increases small-scale power above the
linear growth rate, thereby increasing the noise in the clustering
statistics. This nonlinear process first appears on small scales and
proceeds to larger scales with time. From the perspective of configuration
space, the nonlinear growth damps the BAO peak at $\sim 100\hMpc$ because
the large-scale bulk flows cause the differential motions of the density
pairs initially separated by the sound horizon scale \citep{ESW07}. In
this section, we measure the effect of nonlinear structure growth on
the BAO using the real-space power spectra from our \Nb\ simulations.

\subsection{Fitting Results in Real Space}

\begin{figure}
\plotone{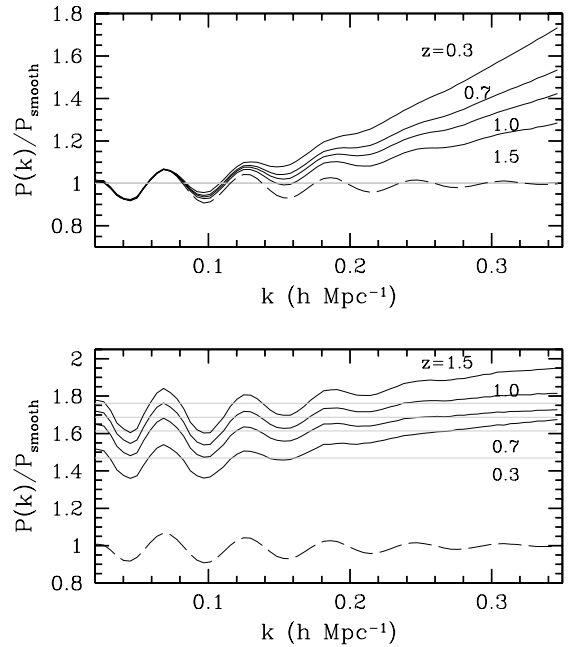}
\caption{Real-space (top) and redshift-space (bottom) power spectrum
$P(k)$ from our $N$-body simulations for z=0.3, 0.7, 1.0, and 1.5. The
power spectra are divided by $P_{\rm smooth}$, the nowiggle form from
\citet{EH98}, scaled with $D^2$. The dashed line is the linear
power spectrum. The gray lines indicate the large-scale amplitude expected
from linear theory. 
\label{fig:power}}
\end{figure}

In the upper panel of Figure \ref{fig:power}, we illustrate the
spherically-averaged real-space power spectra from our \Nb\ simulations
for each of our four redshifts $(0.3$, $0.7$, $1.0$, and $1.5)$, divided
by the nowiggle form from \citet{EH98}. As expected, the effect of
nonlinearity increases the small-scale power and increasingly degrades
the BAO at lower redshifts.

We measure the shifts and the scatters of the shifts of the BAO relative
to linear theory with a $\chi^2$ analysis.  We model the real-space
nonlinear power spectrum by parameterizing $B(k)$ and $\Pano$ in equation
(\ref{eq:Pobs}) in two independent ways.  We use polynomial fitting,
which we refer to as ``Poly7'', with
\begin{eqnarray}
\label{eq:polya7}
\mbox{``Poly7'': }
B(k) &=& b_0 + b_1 k + b_2 k^2 \mathrm{,} \nonumber\\
\Pano &=& a_0 + a_1 k + a_2 k^2 + \mathrm{...} + a_7 k^7 \mathrm{,}
\end{eqnarray}
as well as fitting to Pade approximants, which we refer to as 
``Pade'',  where
\begin{eqnarray}
\label{eq:padec5}
\mbox{``Pade'': }
B(k) &=& b_0 \frac{1 + c_1 k + c_3 k^2 + c_5 k^3}{1 + c_2 k + c_4 k^2}
\mathrm{,} \nonumber\\
\Pano &=& a_0 + a_1 k + a_2 k^2 \mathrm{.}
\end{eqnarray}
Here $a_i$, $b_i$, and $c_i$ are free fitting parameters for all $i$. In
addition, we consider a case in which $\Pano$ in equation (\ref{eq:polya7})
is truncated at third order in $k$, which we 
refer to as ``Poly3''. In constructing $\Pm$, we use reasonable choices
of $\Signl$ that are based on the \Zel\ approximations \citep{ESW07}:
$7.6\hMpc$ at $z=0.3$ and scaling with the linear growth factor at other
redshifts. We vary $\Signl$ near our fiducial values so as to show that
the analysis is not sensitive to the choice of $\Signl$.

We perform a jackknife analysis on our 40 simulations of $P(k)$, using
the two different fitting forms above. We detect non-zero shifts on the
mean acoustic scale [i.e., $\alpha - 1$] of a few tenths of a percent
at all redshifts, as shown in Table \ref{tab:fullreal}. These shifts
represent a decrease in the fitted acoustic scale [i.e., the fitted
BAO being at larger $k$], as opposed to the linear scale, of $0.20 \%$,
$0.26 \%$, $0.33 \%$, and $0.45 \%$ for respective redshifts of $1.5$,
$1.0$, $0.7$, and $0.3$. The shifts increase at lower redshifts, as
expected from the evolution of nonlinear structure. Our results are
statistically significant at the $5 \sigma$, $6 \sigma$, $7 \sigma$,
and $8 \sigma$ levels for respective redshifts of $1.5$, $1.0$, $0.7$,
and $0.3$, for both fitting forms. The reduced $\chi^2$ values in the
table are obtained from the best fit to the averaged power spectrum
over 40 simulations and show the goodness of the fit in general. The
reduced $\chi^2$, $\rechi$, is close to unity even though we have used
a Gaussian approximation for the errors on the nonlinear power spectrum.

\begin{figure}
\plotone{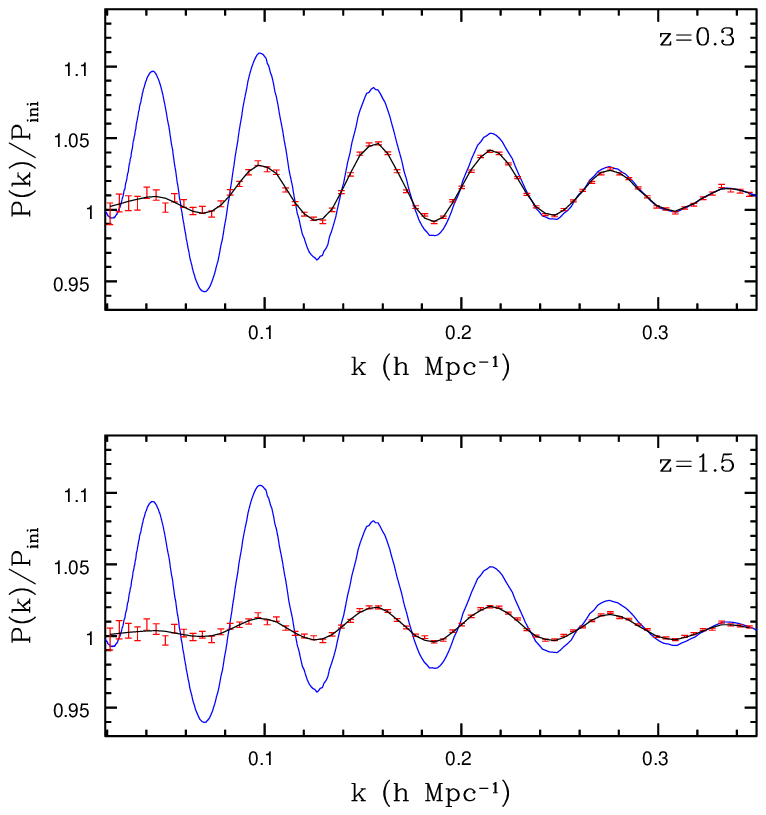}
\caption{Real-space power spectra at $z=0.3$ and 1.5 divided by
the initial linear power spectrum: we divide $\Pres = (\Pnl-\Pano)/B(k)$
by $\Plin$. The oscillatory feature represents the amount of
degradation of the BAO. Red error bars are Gaussian errors centered at
$\Pres/\Plin$. Black solid lines are for the template power spectra $\Pm
(k/\al)/\Plin (k)$ constructed by using $\Signl=7.6\hMpc$ and $4.5\hMpc$
at $z=0.3$ and 1.5, respectively. Blue solid lines represent the nowiggle
form, i.e., $P_{\rm smooth}(k/\al)/\Plin(k)$. 
\label{fig:realfull}}
\end{figure}

\begin{figure*}[t]
\noindent%
\epsfxsize=2.3in\epsffile{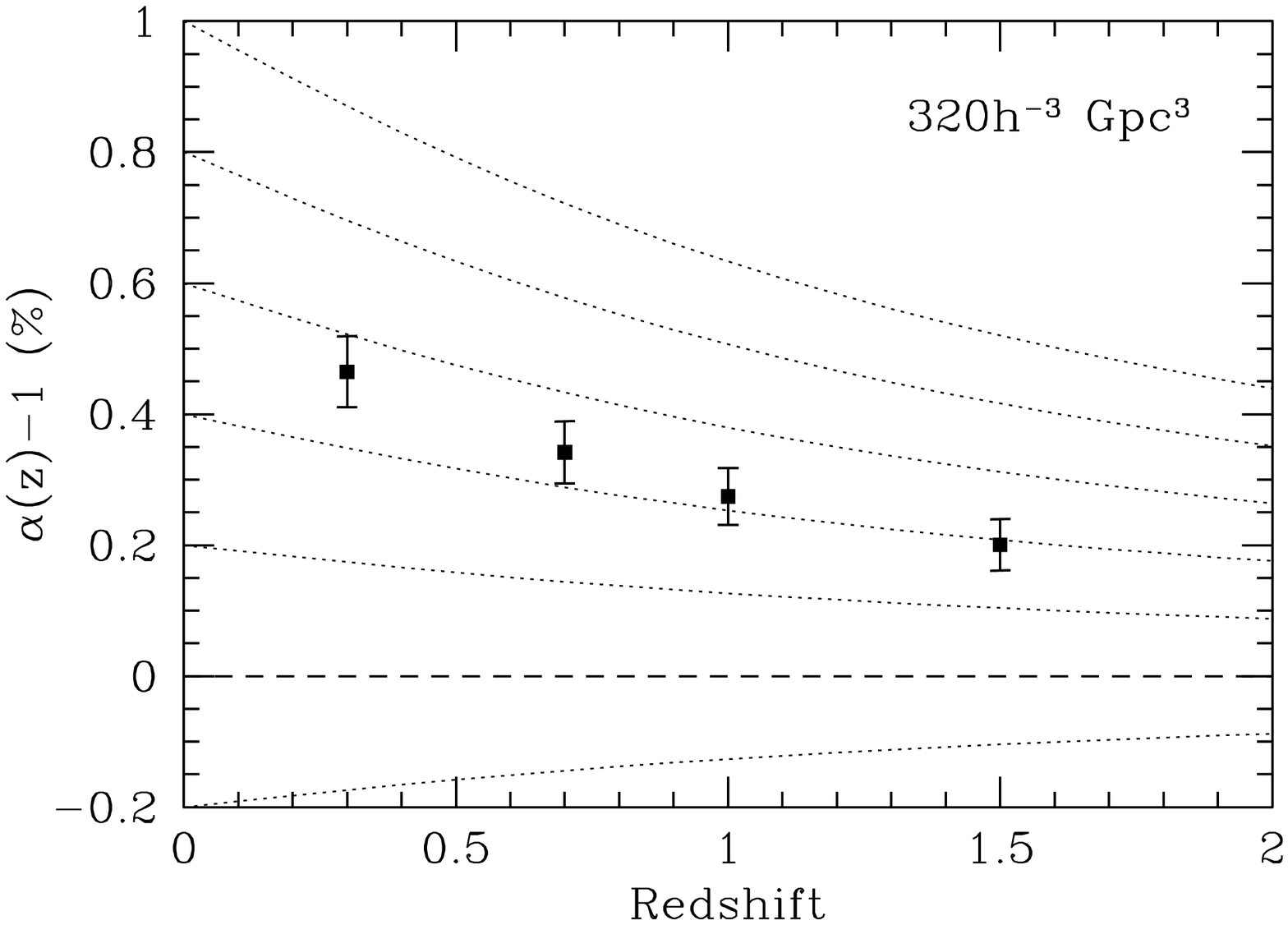}\hfill%
\epsfxsize=2.3in\epsffile{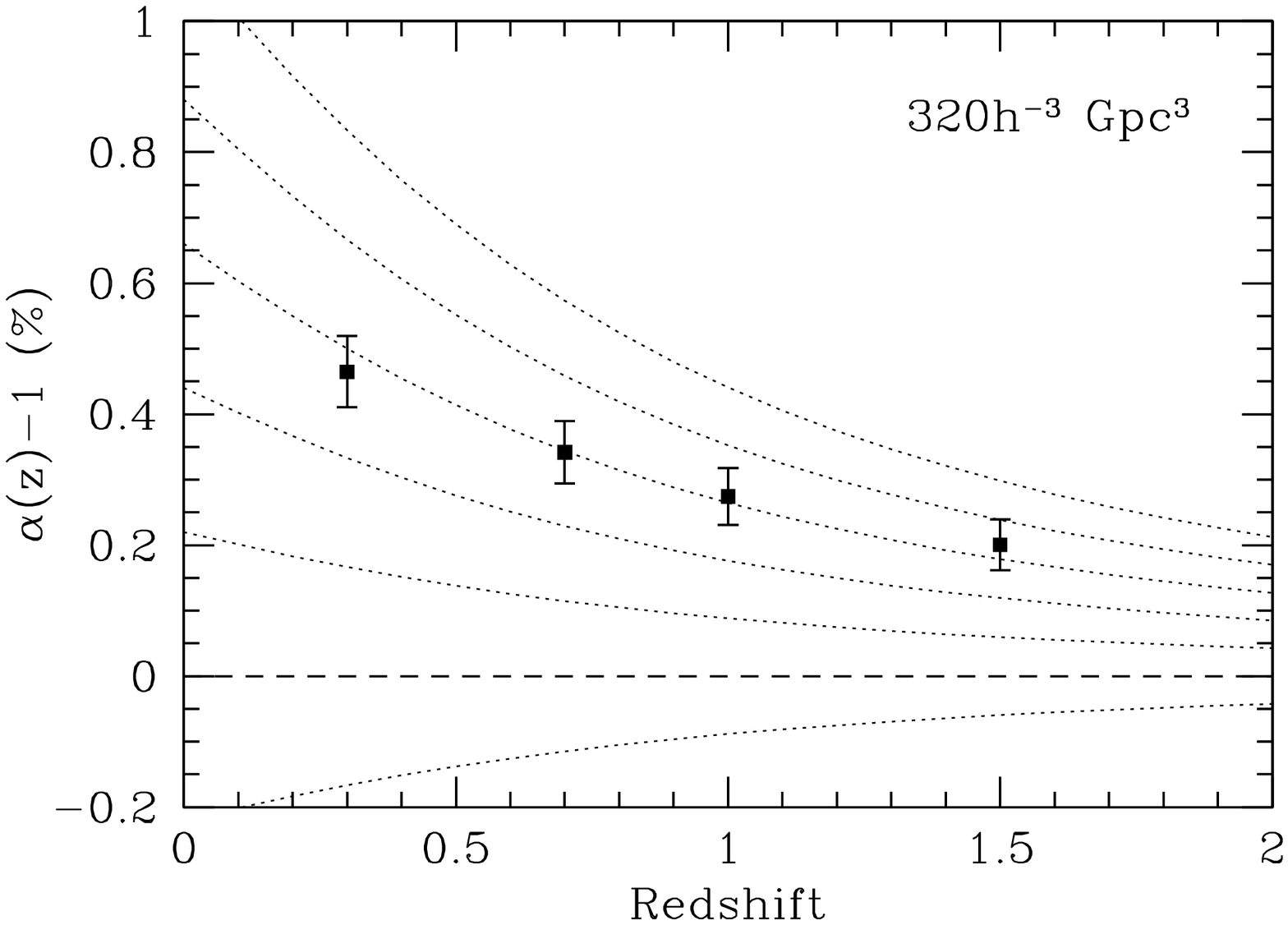}\hfill%
\epsfxsize=2.3in\epsffile{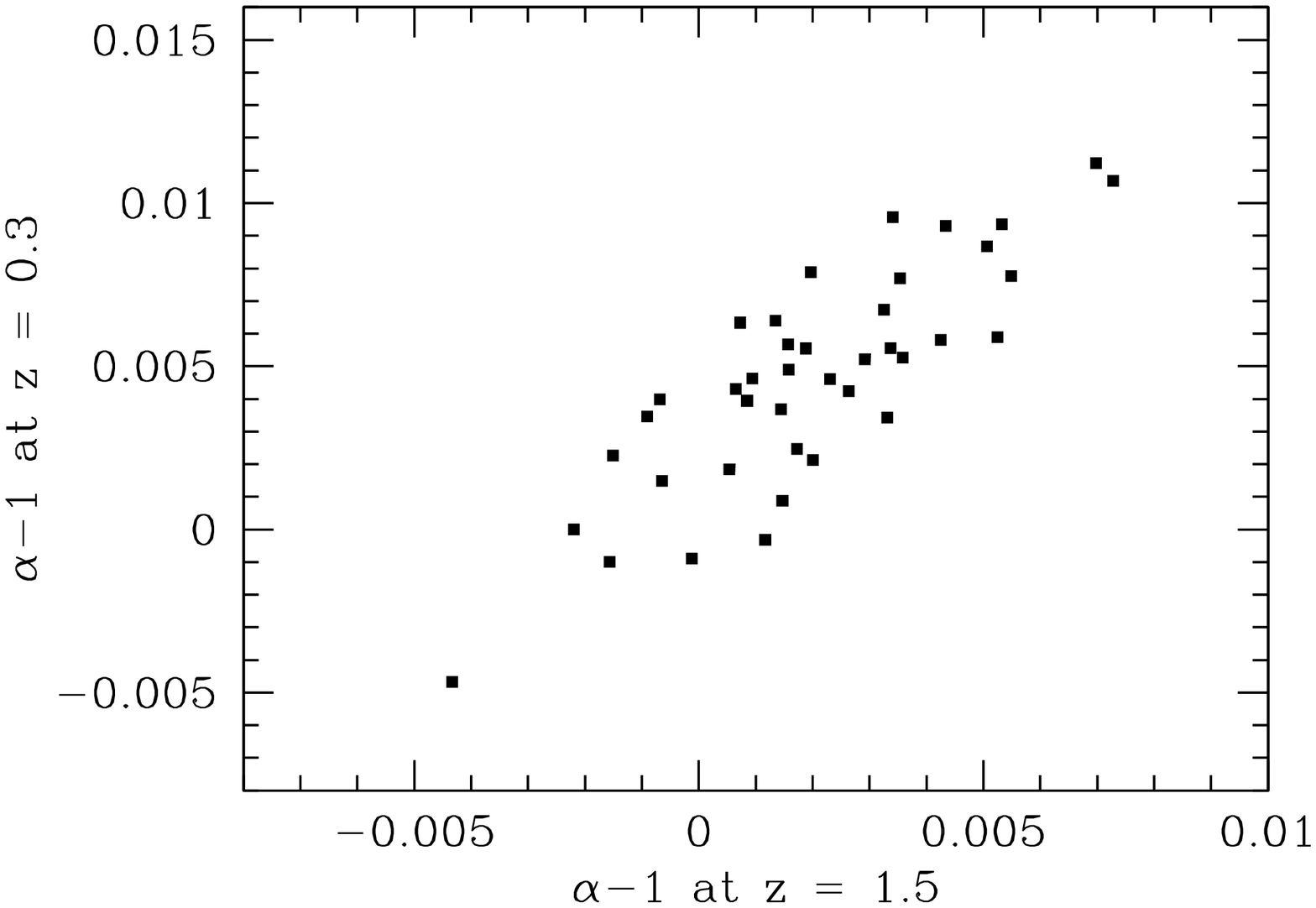}\\
\caption{Left and Middle panels: the growth of nonlinear shift in $\al-1$
with redshift.  The data points show the mean of $\al$ as a function of 
redshift; the errors show the scatter.  The points are correlated.
In the left panel, we overplot curves proportional to
$D^1$; in the middle panel, we overplot curves proportional to $D^2$.
Right panel: the $\al-1$ at $z=0.3$ and $z=1.5$ for 40 simulations.
One can see that a given simulation produces $\al$ that are highly 
correlated between redshifts.
\label{fig:growthal}}
\end{figure*}

Our results are consistent over a wide range of $\Signl$ ($\Delta
\Signl=\pm 2\hMpc$) used in $\Pm$, as shown in the lower half of Table
\ref{tab:fullreal}: the difference in $\al$ is well within $\sial$ and
the difference in $\sial$ is well within the expected Gaussian sample
variance of $1/\sqrt{2\times 40}\approx 11\%$ for 40 simulations. For example,
varying $\Signl$ between $6.0\hMpc$ and $9.0\hMpc$ at $z=0.3$ causes 
negligible changes in $\al$ and $\sial$ (less than $10^{-4}$ in $\al$).
Reasonable variations at other redshifts are similarly negligible.
That is, due to the
large number of parameters allowed in our parameterization, our fitting
is able to tolerate a mild discrepancy between $\Signl$ used in $\Pm$
and the actual power spectrum, without a substantial degradation of the fit.

While the ``Pade'' model gives a slightly better fit than the ``Poly7''
model, based on the smaller $\rechi$ values, we note that both models
give the same acoustic peak scale and very similar errors. 
We tried varying the order of the fitting formulae and found that our
results did not change.
Of course, one cannot make $B(k)$ and $\Pano$
arbitrarily flexible or else these terms would be able to mimic the BAO. With a fitting range
of $0.02\ihMpc \leq k \leq 0.35\ihMpc$, the flexibility we allowed in
both our fitting formula avoids this problem.  As shown in Table
\ref{tab:fullreal}, decreasing the polynomial orders of $\Pano$ to the
third order, i.e., ``Poly3'',  makes a negligible change to the best-fit
$\al$ and $\sial$.  However, using a model in which $B(k)$ is constant
produced notably worse fits and made the fit more sensitive to the choice of
$\Signl$ used in $\Pm$. To be conservative, we recommend a second order
$B(k)$ and higher than fourth order in $\Pano$ in real space.

We  find that the bootstrap analysis is highly consistent with the
jackknife estimate of the scatter in $\al$. Both agree to within 5\%
in $\sial$ and $0.0001$ in $\al$.

Figure \ref{fig:realfull} shows the real-space average power spectrum at
$z=1.5$ and 0.3, and, as a comparison, the best-fit models to them. Here,
we first subtract the additive part $\Pano$ from the nonlinear power
spectra and then divide the residual $\Pnl-\Pano$ by $B(k)$ and
by the input linear power spectrum, so as to manifest the degradation
in the BAO feature. That is, the broad-band differences, relative to the
initial power spectrum, have been removed and only the difference in the
BAO structure is illustrated: any oscillations in the figure represents
the nonlinear {\it erasure} of the BAO. We note that the BAO feature in
Figure \ref{fig:realfull} depends on the value of $\Signl$ we assume for the
template $\Pm$. The results at $z=0.7$ and 1 are similar to
the redshifts shown.  Error bars are drawn assuming
Gaussian errors, which will be an underestimation on small scales, where
there should be a contribution from the trispectrum. We also ignore the
nonlinear covariance between the different data points in the figure. The
excellent agreement between the best-fit power spectrum and the \Nb\ data 
demonstrates that our template model has a suitable amount of flexibility to
match the data well.

To conclude, we find shifts of 0.2\% to 0.45\% in the acoustic scale
due to nonlinear growth. The shifts increase at lower redshifts. 
Our recovery of the acoustic scale from the 
template fitting method is robust for different fitting forms, a wide
range of $\Signl$, and the choice of jackknife or bootstrap resampling.

\subsection{The Growth  of $\al-1$ with Redshift}
We next investigate the redshift dependence of the shift $\al$.
The left and the middle panel of Figure \ref{fig:growthal}
shows the growth of $\al-1$ with redshift with the error bars $\sial$
for $320\trihGpc$. The dotted lines are curves of linear growth with
different proportionality constants in the left panel, and are curves
scaling quadratically with the growth factor in the middle panel. From
the figure, $\al-1$ is growing faster than the growth factor, $D$,
but slower than $D^2$. We find that $\al-1$ scales as $D^m$ where
$m= 1.66^{+0.26}_{-0.22}$, when we perform a $\chi^2$ fitting to the
mean $\al$ with a covariance matrix generated by the variations in
the bootstrap samples. The right panel shows that values of $\al-1$
for each subsample are highly correlated between different redshifts,
as expected. A model of $\al-1$ scaling as $D^2$, which one might
expect from the perturbation theory \citep{Crocce08}, is only $1.3\sig$
away from the best fit; we cannot reject this model. Meanwhile, $\al-1$
scaling as $D$ is disfavored at about $3\sig$.  
However, a model with leading order
$D$ and higher-order terms fits well and cannot be excluded.

We next compare our results with the analytic predictions. While we
detect the sub-percent level nonlinear mode-coupling shift predicted by
\citet{Smith08} and \citet{Crocce08} \citep[also see][]{Nishimichi07},
the redshift-dependence of $\al-1$ we derive is weaker than the
node-by-node shifts in the BAO feature in power spectrum predicted
by \citet{Crocce08}, where they simulate the fitting method used in
\citet{Percival07b} \footnote{Their $\al$ corresponds to the inverse
of $\al$ defined in this paper.}. Note, however, that we are using a
global fit, where the overall shift will be a result of different nodes
being weighed differently depending on redshift, resulting in an overall
redshift dependence that is not the same as the redshift dependence of
individual nodes. Indeed, when they consider a global fit, the redshift
dependence of the shift seems attenuated [Eq. (27) of \citet{Crocce08}].
In addition, other details of our fitting method are different from the
cases they considered. Taking these differences into account, we consider
that these analytic works and our \Nb\ results are largely in agreement.

\section{A redshift-space analysis of the BAO}\label{sec:reddis}
\subsection{Redshift-space Distortions}

\begin{figure}
\plotone{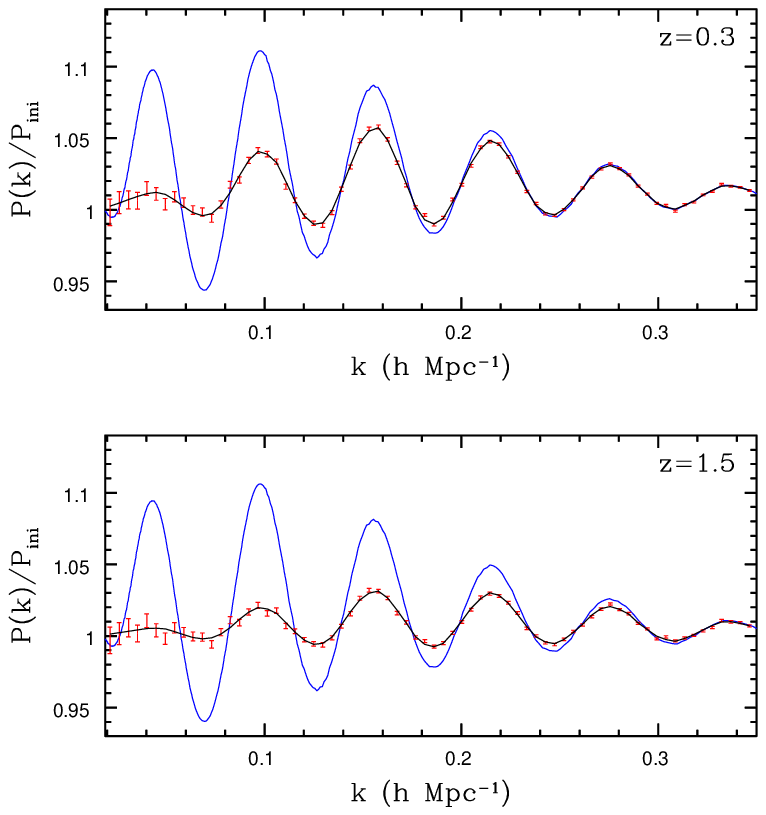}
\caption{Redshift-space power spectra at $z=0.3$ and 1.5 divided by the
initial linear power spectrum: we divide $\Pres = (\Pnl-\Pano)/[B(k)\FoG]$
by $\Plin$. The oscillatory feature represents the amount
of degradation of the BAO. Red error bars are Gaussian errors
centered at $\Pres/\Plin$. Black solid lines: $\Pm (k/\al)/\Plin (k)$
constructed by using $\Signl=9.0\hMpc$ and $6.0\hMpc$ at $z=0.3$ and
1.5, respectively. Blue solid lines represent the nowiggle form, i.e.,
$P_{\rm smooth}(k/\al)/\Plin(k)$.
\label{fig:redfull}}
\end{figure}

The observed power spectrum is subject to redshift-space distortions
due to the peculiar velocity of the observed sources. The large-scale
bulk motions [i.e., outflow from underdense regions and infall
toward overdense regions] along the line of sight increase
the apparent density contrast in this direction and induce an
angle-dependent amplification of the real-space power spectrum on
large scales \citep[][etc]{Kaiser87,Hamilton98,Sco04,Papai08}. The
apparent displacements due to such redshift distortions
degrade the BAO further in redshift space relative to real space
\citep[e.g.,][]{Meiksin99,SE05,ESW07}.  Meanwhile, virial motions between
and within halos cause an apparent suppression of the small-scale power
along the line of sight, which is called the finger-of-God effect (FoG)
\citep{deLa86}.

While the redshift distortions are an inevitable element of the
study of cosmological distance error estimation, a direct $\chi^2$
analysis of the redshift-space power spectrum has been challenging
due to the lack of a reliable model for fitting redshift distortions
and the anisotropic nature of the redshift-space power spectrum
\citep[cf.,][]{Wagner07,Angulo08,Okumura08,Pad08}. Recent theoretical and numerical
studies \citep{ESW07,Mat08}, however, provide quantitative predictions
for the BAO degradation due to nonlinearity and redshift distortions and
therefore for the resulting distance errors \citep[e.g.,][]{SE07}. Based
on these studies, we can improve the description of the nonlinear effect
on BAO in the fitting model and check the distance errors derived from
the redshift-space power spectra.

In this section, we extend the $\chi^2$ analysis to the
spherically-averaged redshift-space power spectrum to find the effect
of the redshift distortions on the acoustic scale measurements. 
We do not include large-angle effects in the redshift distortions but
merely use the flat-sky approximation.  
Due to a computer hardware problem, some of the redshift-space data were lost,
so that our redshift-space analysis uses a volume of $216\trihGpc$ (i.e.,
27 simulations).

\subsection{Fitting Formulae}\label{subsec:redfitting}
The lower panel of Figure \ref{fig:power} shows the spherically averaged
redshift-space power spectra from our simulations. In order to account
for the FoG effect in a $\chi^2$ analysis, we add a simple multiplicative
function $\FoG$ to our parameterization. We try two ways of incorporating
$\FoG$ into our parameterization:
\begin{eqnarray}\label{eq:redin}
\mbox{``In'':  } 
P(k)&=& B(k) P_m(k/\al) \FoG +\Pano, 
\end{eqnarray}
and 
\begin{eqnarray}\label{eq:redout}
\mbox{``Out'':  }
P(k)&=&[ B(k)P_m(k/\al) + \Pano] \times \FoG  \nonumber\\
&+& e_1.
\end{eqnarray}
where $\Pano=a_0 + a_1 k + a_2 k^2$. Note that the highest order of
polynomials for $\Pano$ is only $k^2$, less than that of ``Poly7'' in real
space. We lower the number of parameters in $\Pano$ in redshift space,
as there are enough additional parameters to account for the broad-band
shape. The parameter $e_1$ is added to permit a shot noise term in the
``Out'' model .

We again try both polynomial and Pade approximants for $B(k)$:
\begin{eqnarray}
\mbox{``Poly'': }
B(k)&=&b_0+b_1 k+b_2 k^2  \label{eq:bkpoly}
\end{eqnarray}
and
\begin{eqnarray}
\mbox{``Pade'': }
 B(k)&=&b_0\frac{1+c_1 k+c_3 k^2 + c_5 k^3}{1+c_2 k+c_4 k^2}  \label{eq:bkpade}.
\end{eqnarray}
We try two forms for $\FoG$: 
\begin{eqnarray}\label{eq:invpow}
\mbox{``Rat'': }
\FoG&=&1/\left[1+(k \don)^{\dt}\right] ^{1/\dt} 
\end{eqnarray}
with  priors of $\don >0.1 \hMpc$ and $\dt>0$, and 
\begin{eqnarray}\label{eq:exp}
\mbox{``Exp'': }
\FoG&=&\exp\left[-(k \don)^{\dt}\right] 
\end{eqnarray}
with priors of $\don > 0\hMpc$ and $\dt>0$. We refer to, for example,
the combination of equations (\ref{eq:bkpoly}) and (\ref{eq:exp}) in the
form of equation (\ref{eq:redout}) as ``Poly-Exp-Out''. In addition, we
also try ``Poly7'' (eq. [\ref{eq:polya7}]) to fit the redshift-space power
spectrum. We will mainly quote the results from using ``Poly-Exp-Out''.

We increase the values of $\Signl$ used in $\Pm$ relative to the
choices in real space, so as to account for the additional degradation
of the BAO due to redshift distortions. Note that we are approximating
the anisotropic nonlinear effect on the BAO in redshift space with an
isotropic parameter $\Signl$.

\subsection{Fitting Results in Redshift Space}\label{subsec:Resultsred}
Table \ref{tab:fullred} lists the errors and the mean of the best-fit
$\al$ of jackknife samples in redshift space. We also present the values of $\sial$
rescaled to a survey volume of $320\trihGpc$ from the actual
results for $216\trihGpc$ so as to provide an easy comparison to Table
\ref{tab:fullreal}.

We again detect a non-zero, sub-percent shift of the mean values of $\al$
in redshift space: 0.25\% of shift of the acoustic scale at $z=1.5$
by $5\sig$, 0.33\% at $z=1$ by $5\sig$, 0.41\% at $z=0.7$ by $6\sig$,
and 0.54\% at $z=0.3$ by $6\sig$, where the signal-to-noise ratios are
calculated based on the original $\sial$ for $216\trihGpc$. The shift
on $\al$ has systematically increased relative to the real-space values
at all redshifts. We find that $\al$ increases by $\sim25\%$ at all
redshifts; we do not detect a redshift-dependence in this increase. As
expected, values of $\sial$ also have increased relative to the real-space
values at all redshifts: about 25\% at $z=0.3$ and 0.7, 18\% at $z=1$,
and 11\% at $z=1.5$.

We find our results to be consistent for both $\al$ and $\sial$,
regardless of the various parameterizations.  As shown in Table
\ref{tab:fullred} for $z=0.3$, the differences in $\al$ are well within
$\sial$ and the differences in $\sial$ are well within the sample variance,
$1/\sqrt{2\times 27} \approx 14\%$, in most of the cases. Interestingly, we find that even
without correcting for $\FoG$, such as the real-space fitting formula
``Poly7'' (eq [\ref{eq:polya7}]), yields results consistent with ones
with $\FoG$ correction, so long as enough free fitting parameters
are allowed. Again, the results are not very sensitive to the fiducial
values of $\Signl$ used to calculate $\Pm$. We therefore choose equation
(\ref{eq:redout}) combined with the polynomial $B(k)$ and the exponential
form of $\FoG$ (i.e., eq [\ref{eq:exp}]), ``Poly-Exp-Out'', as our
convention to fit the redshift-space power spectra for the rest of paper.

Figure \ref{fig:redfull} shows the redshift-space power spectrum at
$z=1.5$ and 0.3 and the best-fit models to them. Again,
we first subtract the additive part $\Pano$ from the nonlinear power
spectra, but this time we divide the residual $\Pnl-\Pano$ by
$B(k)\FoG$ before dividing by the input linear power spectrum, so as to
manifest the degradation in the BAO feature.  The figure illustrates the
excellent agreement between the best-fit power spectrum and the \Nb\ data.

\section{Reconstructing the BAO}\label{sec:recons}
\begin{figure*}
\plotone{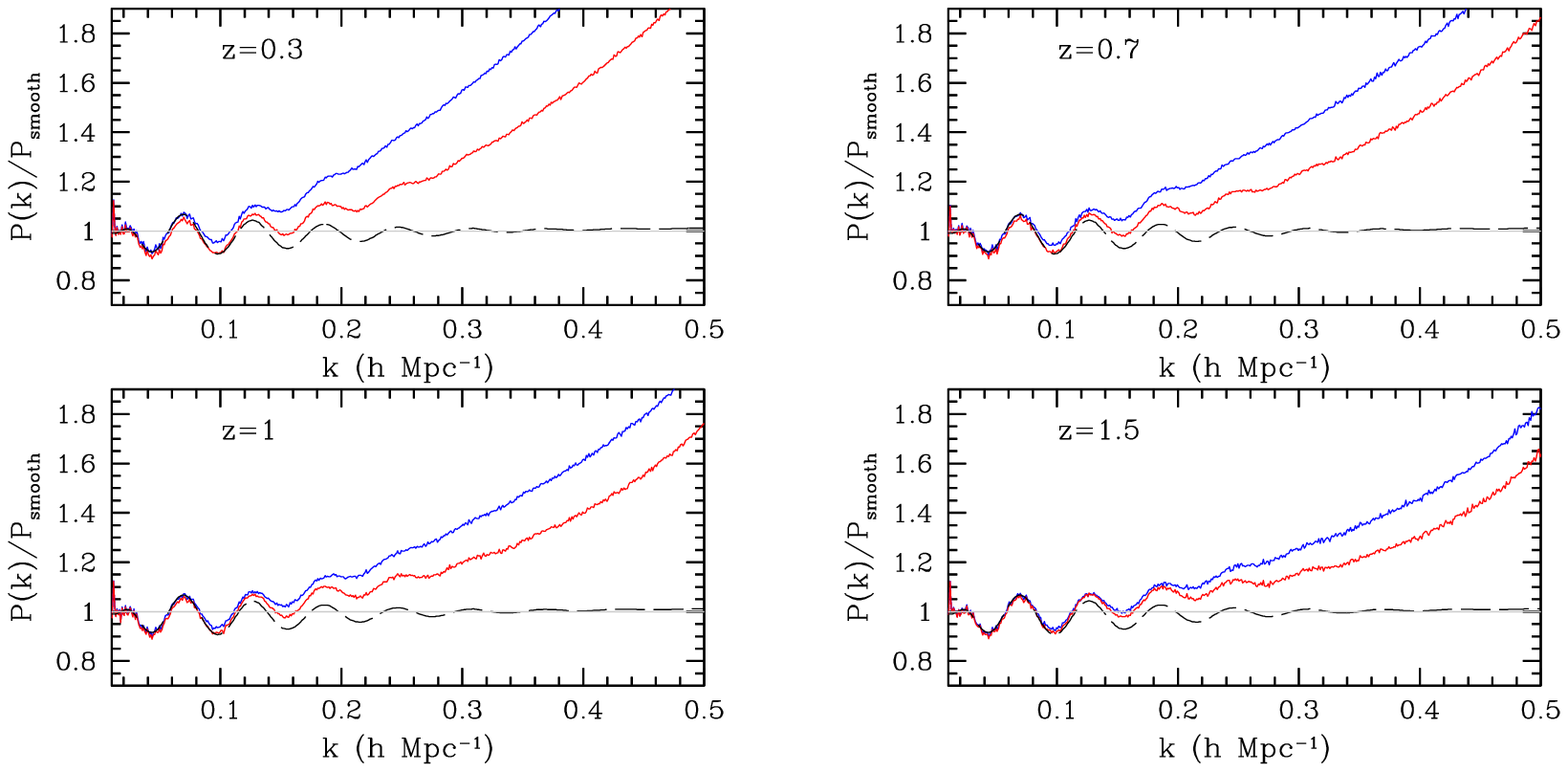}
\caption{Real-space $\Pnl(k)$ of the 1\%-sample after reconstruction (red
lines), divided by the nowiggle form from \citet{EH98}, in comparison to
the nonlinear power spectra before reconstruction (blue lines). The gray
lines are for the large-scale amplitude expected from linear theory. The
dashed lines are for linear power spectrum. The effect of reconstruction
is most apparent at $z=0.3$ and is smallest at $z=1.5$.
\label{fig:dlnPdlnkreal}}
\end{figure*}

\begin{figure}[t]
\plotone{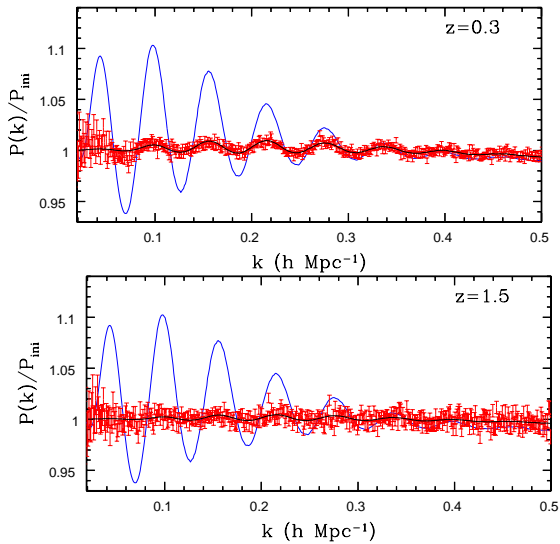}
\caption{Real-space $\Pnl(k)$ of the 1\%-sample after reconstruction
(red error bars): we divide $\Pres = (\Pnl-\Pano)/B(k)$ by $\Plin$. The
oscillatory feature represents the amount of degradation of the BAO. Red
error bars are Gaussian errors centered at $\Pres/\Plin$. Black solid
lines are for the template power spectra $\Pm (k/\al)/\Plin (k)$
constructed by using $\Signl$ listed in Table \ref{tab:recreal}. Blue
solid lines: $P_{\rm smooth}(k/\al)/\Plin(k)$.
\label{fig:recreal}}
\end{figure}

In this section, we reconstruct the baryon acoustic signature obscured
by nonlinear growth and redshift distortions using the simple scheme
presented in \citet{ESSS07}.  We wish to investigate whether 
such reconstruction improves the signal-to-noise ratio on the 
measurement of $\al$ \citep{SE07} and whether
it reduces the shift of the acoustic scale.

For the analysis with reconstruction, we use the phase-space data of a randomly
selected 1\% of all the particles. We will refer to this sample as
``1\%-sample'', in comparison to the previous ``100\%'' sample. For
the 1\% sample, power spectra are computed for wavenumber bins of width
$\Delta k=0.001\ihMpc$ and we use a fitting range of $0.02\ihMpc \leq
k \leq 0.5\ihMpc$. We utilize a volume of 320$\trihGpc$ both in real
space and in redshift space. We note that by using the 1\% sample,
we perform the reconstruction scheme in the presence of non-negligible
shot noise.  \citet{SE07} quantify the ratio of shot noise to sample 
variance with $\nPt$, where $n$ is the number density and $P_{0.2}$
is the real-space power at $k=0.2\ihMpc$.  For the analysis here,
$\nPt$ runs from 0.7 at $z=1.5$ to 2.4 at $z=0.3$. 

We apply the linear perturbation theory continuity equation to the nonlinear 
density fields
of the 1\%-sample and predict the linear theory motions. We first take the
nonlinear density fields at z=0.3, 0.7, 1.0, and 1.5 and Fourier-transform
them. We apply a $10\hMpc$ Gaussian filter to $\hat{\delta} (k)$ in order
to smooth gravity on small scales. We then compute the linear theory
motions $\vec{q}$ using $\bigtriangledown \cdot \vec{q}=-\delta$. Finally
we displace the real particles and smoothly distributed reference
particles by the resulting $\vec{q}$, and find the final density field
from the difference of the two density fields. As \citet{ESSS07} noted,
this differencing method moves the measured {\it densities} back to their
initial locations: linear structure growth is mostly preserved and only
the nonlinear growth effect is reduced as a result.

\subsection{Reconstruction in Real Space}

Figure \ref{fig:dlnPdlnkreal} shows the reconstructed power spectra of
the 1\% sample, divided by $P_{\rm smooth}$, at various redshifts. The
benefits of reconstruction are most obvious at $z=0.3$ due to the low
shot noise and the strong nonlinearity to correct for at this redshift.
We fit the power spectra of these reconstructed density fields using the
same models as before (eq.\ [\ref{eq:polya7}] and [\ref{eq:padec5}]). 
We find consistent results for both models in general. 
Because the upper limit of the fitting range was chosen to be $k=0.5\ihMpc$
in the 1\% case, we find that the ``Poly7''
model with more flexible $\Pano$ fits better to the reconstructed
real-space power.
We decrease $\Signl$ for the template $\Pm$ approximately by half
relative to $\Signl$ values used before reconstruction.

Table \ref{tab:recreal} shows the resulting $\al-1$ and $\sial$
for the ``Poly7'' model.  Because the
reconstruction analysis uses the 1\% sample, the results in Table
\ref{tab:recreal} are not immediately comparable to those in Table
\ref{tab:fullreal}, which used the 100\% sample and hence had negligible
shot noise. We include results for the 1\% sample without reconstruction
in Table \ref{tab:recreal} to provide the proper comparison. 

We find that applying reconstruction decreases $\sial$ by almost 
a factor of 2 at $z=0.3$. 
The improvement is incrementally less at higher redshifts.
This is to be expected for several reasons.
First, the density field is more linear at high redshift.  As most of
the acoustic scale information is at $k\lesssim 0.2\ihMpc$, 
the non-linear degradation of $\sial$ is smaller at high redshift.  
There is therefore less for the reconstruction algorithm to fix.
Second, the intrinsic clustering is weaker at high redshift such that
the shot noise is more important.  This contaminates the measurement
of the density field and hence the estimation of the reconstruction
displacements, causing the reconstruction to be less accurate.
Third, the increased shot noise means that the high harmonics are
not being well measured, regardless of whether or not they are correctly 
reconstructed.
This means that the results with and without reconstruction will not
differ much.
For example, the comparison of Tables \ref{tab:fullreal} and
\ref{tab:recreal} (summarized in the ``N-body'' column of Table 
\ref{tab:Fisher}) shows that the introduction
of shot noise degrades $\sial$ by over a factor of 2 at $z=1.5$,
even without reconstruction.
Finally, the force resolution of our PM simulations ($1\hMpc$ mesh) may not be 
adequate to recover the BAO with full fidelity when our reconstructed
$\Signl$ is expected to be as small as $2-3\hMpc$, as it is at $z>1$.
If our small-scale structure is more fluffy, this will appear as an
extra source of $\Signl$.

In addition to the improvement in the scatter of $\al$, 
Table \ref{tab:recreal} also shows that reconstruction 
decreases the shifts in $\al$ itself. The shift in
$\al$ decreases to 0.07\% at $z=0.3$ and to below 0.03\% at higher
redshifts. Note that while reconstruction does not
improve $\sial$ at $z=1.5$,  
it does substantially improve the shift in $\al$. 
This likely indicates that the reconstruction method is correctly 
compensating for the large-scale bulk flows but is leaving a small-scale
jitter that washes out the high harmonics of the BAO, as discussed above.
Figure \ref{fig:recreal}
shows the best-fit models to the averaged power spectra of the 1\%-sample
after reconstruction. Increasing the Gaussian filter size to $14\hMpc$
makes no difference: less than 2\% in $\sial$ and $10^{-4}$ in $\al$.

\subsection{Reconstruction in Redshift Space}

\begin{figure*}[t]
\plotone{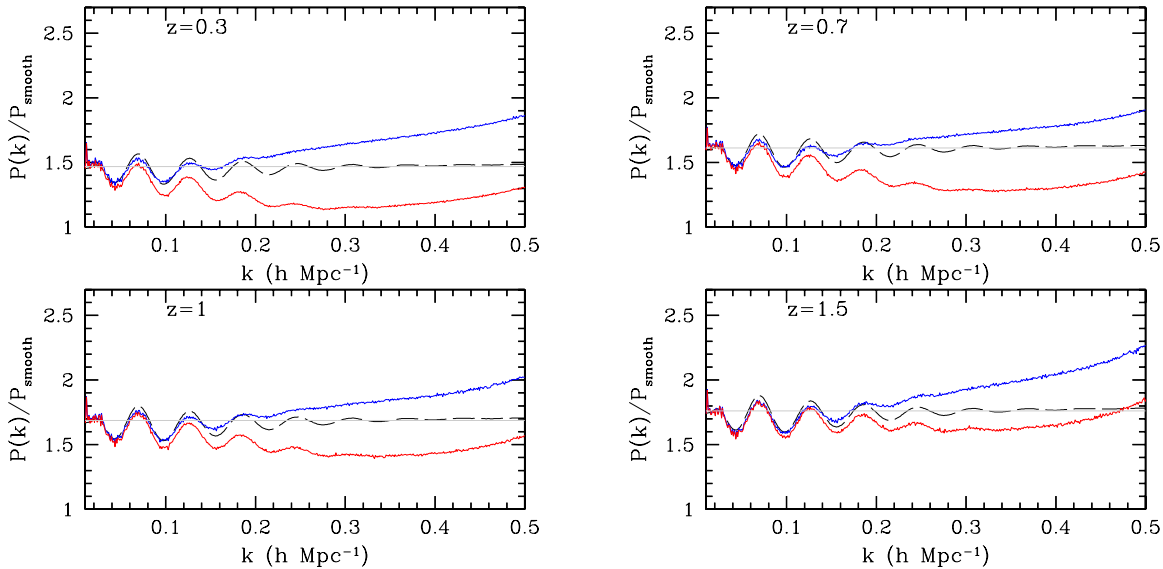}
\caption{Redshift-space $P(k)$ of the 1\%-sample after reconstruction (red
lines), divided by the nowiggle form, in comparison to the redshift-space
nonlinear power spectra before reconstruction (blue lines). The gray lines
are for the large-scale amplitude expected from linear theory. The dashed
lines are for linear power spectrum.   The effect of reconstruction is
most apparent at $z=0.3$ and is smallest at $z=1.5$}
\label{fig:dlnPdlnkred}
\end{figure*}

\begin{figure}[t]
\plotone{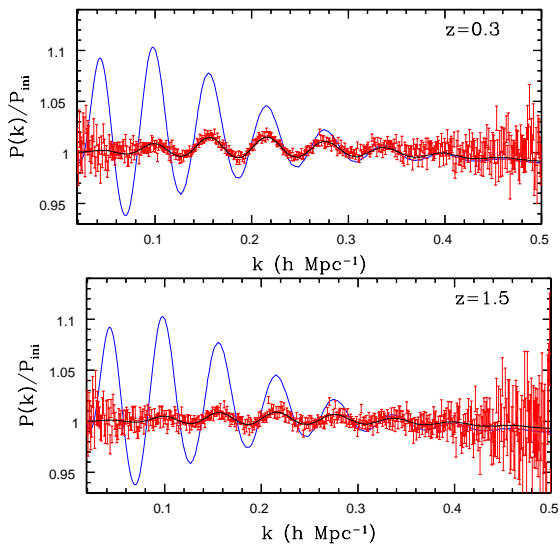}
\caption{Redshift-space power spectra of the 1\%-sample after
reconstruction (red error bars), in comparison to the best-fit models
(black lines) and nowiggle form (blue lines): we divide $\Pres =
(\Pnl-\Pano)/B(k)/\FoG$ by $\Plin$. }
\label{fig:recred}
\end{figure}

Next we apply the reconstruction techniques to the redshift-space density
field.  The method is unchanged, save that we use the observed nonlinear 
redshift-space density field and apply the resulting displacement field to
the redshift-space positions \citep{ESSS07}.  In this paper,
we do not attempt to correct for the FoG effect of virialized halos
\citep[c.f.,][]{ESSS07}.

Figure \ref{fig:dlnPdlnkred} shows the redshift-space power spectra
after reconstruction, divided by the nowiggle form; the difference
before and after reconstruction is significant even in redshift space,
not only at $z=0.3$ but also at $z=0.7$ and 1.0.  Table \ref{tab:recred}
quantifies the values of $\al$ and its scatter after applying the
reconstruction method in redshift space. We find that the scatter $\sial$
is substantially reduced to a level comparable to that found with the
reconstructed real-space case. That is, the reconstruction method seems more
effective in redshift space.  The BAO is more smeared in redshift space, 
providing more opportunity for reconstruction to improve $\sial$.
In addition, the higher
amplitude of the power spectra in redshift space reduces the effects of
shot noise and may permit better reconstruction.

As in real space,
the shifts in the acoustic scale, $\al$, are markedly improved, decreasing to 
below 0.1\% at all redshifts. Figure
\ref{fig:recred} shows the resulting best-fit to the averaged power
spectra, compared to the data. Again, we find a good agreement between
the two.

\section{Comparing the N-body Results to the Fisher Matrix Prediction}\label{sec:fisher}
We next compare $\sial$ from the \Nb\ data with the analytic distance
error estimates from the Fisher matrix formalism.  We calculate the Fisher
matrix estimates by using a fitting formula based on \citet{SE07}. 
This formula uses a parameter $\Signn$ to quantify the amount of non-linear
degradation of the BAO.  
We estimate a
reasonable value of $\Signn$ at each redshift by using the analytic \Zel\
approximations (eq. [9] of \citet{ESW07}).

Table \ref{tab:Fisher} compares $\sial$ from the N-body data with
the Fisher matrix estimates. We start with the Fisher matrix errors
corresponding to the 100\% samples, for which shot noise is negligible; we
use $\nPt=100$ for all redshifts. We do not include the nonlinear growth
effect into $\nPt$, following the convention in \citet{SE07}. While the
Fisher matrix errors are in reasonable agreement with the N-body data, the
Fisher matrix calculation systematically overestimates $\sial$ by $\sim
25\%, 17\%$, 15\%, and 5\% at $z=0.3$, 0.7, 1.0, and 1.5, respectively,
as shown in Table \ref{tab:Fisher}. Such an overestimation is contrary to
our intuition that the \Nb\ data would reflect extra nonlinear effects,
such as an effect on variance of the power spectrum, that the Fisher
matrix does not incorporate. As we allow many free parameters in our
fitting forms, the contribution to the standard ruler method from the
broad-band shape of the power spectrum is negligible. However, because
$\sial$ is estimated from only 40 simulations, the error on the scatter
is expected to be $1/ \sqrt{2\times 40}\approx 11\%$, assuming Gaussianity. While the discrepancy becomes larger than the expected sample variance at $z\leq 1$, it is possible that the N-body results are lower than the Fisher matrix
predictions simply because of sample variance. Because $\al$ values in
a simulation are largely correlated between redshifts, it would not be
surprising that the residuals in $\sial$ could all have the same sign.

For the 1\%-samples (Table \ref{tab:Fisher}), shot noise is not
negligible. We derive $\nPt$ from the nominal shot noise, i.e., the
inverse of the number density of particles. We find that the Fisher matrix
errors are bigger than those from the simulations by 18\% and 23\% at 
$z=0.3$ and 0.7, respectively. At
$z=1.0$ and 1.5, the agreement is better.  The degradation in $\sial$ 
from the simulation due to the addition of shot noise appears well
predicted by the Fisher matrix formula.  We note that the shot noise
and its effect on $\al$ is not correlated between redshifts, leading 
to a little more scatter around the redshift trends.

We next compare $\sial$ from the \Nb\ data with the Fisher matrix errors
in redshift space. We use the 2-dimensional [2-D] approximation based
on \citet{SE07} and contract the 2-D matrix to derive the 1-D distance
error, as the transverse and the line-of-sight distance scales are the
same in this simulated Universe. We use an ellipsoidal smoothing kernel
with $\Sigxy$ ($=\Signn$) and $\Sigz=[1+\Om(z)^{0.6}]\Sigxy$ and include
the angle dependence in $\nPt$. The FoG effect is not included in $\nPt$,
as justified in \citet{SE07}.

From Table \ref{tab:Fisher},  the $\sial$ values from the 100\%-samples
deviate from the Fisher matrix errors approximately by 1.5 times the sample
variance. Again, they are systematically lower than the Fisher matrix
errors. Note that we expect the sample variance of $1/ \sqrt{2\times 27}  \approx
14\%$ rather than $1/ \sqrt{2\times 40}$ in this case. For the 1\%-samples, the
discrepancy reaches up to 27\% at $z=0.3$, while  the discrepancy is 
below the sample variance of $1/ \sqrt{2\times 40}  \approx 11\%$ at high redshift.

Overall we find that the distance errors from the \Nb\ data in real and
redshift space
are somewhat smaller than the Fisher matrix errors with $\Signn$ based
on the \Zel\ approximations.  The measured errors from the \Nb\
data correspond to reductions of $\Signn$ by as much as $1\hMpc$
relative to the \Zel\ prediction. 
It is possible that this systematic discrepancy is merely due to sample 
variance, which is expected to be correlated between redshifts.
However, there could be other reasons for the difference.  It is possible that the \Zel\ 
approximation slightly overestimates the non-linear degradation \citep[][Padmanabhan \& White, in preparation]{Crocce06b} or that
the approximations behind the Fisher matrix calculation are slightly 
pessimistic.  Previous comparisons to high-resolution N-body simulations
\citep{SE07} found agreement to better than 13\%, comparable to the 
precision tested here.  In principle, we could try to use the amplitude
of the wiggles in $P(k)$ to estimate $\Signn$, but our template fitting
parameter $\Signl$ is not a precise proxy for $\Signn$ because the 
marginalization parameters allow good fits for a range of $\Signl$.

In conclusion, we are encouraged by the good agreement between the 
N-body and Fisher matrix results.
The fact that the $\sial$ from the N-body fits are not much worse than
those from the Fisher matrix indicates that 
our estimator is extracting most of the acoustic information.

\section{Effects of Variations in the Template Cosmology}\label{sec:robust}
When the true acoustic scale is known, the observations of the BAO
allow us to measure the cosmological distance scale, in particular
the angular diameter distance $D_A(z)$ and the Hubble parameter $H(z)$.
The true acoustic scale is determined from measurements of the matter
and baryon density, e.g., from CMB anisotropy data.  In this paper,
we do not marginalize over uncertainties in the cosmological densities
but instead use a fixed fiducial template.
Using a fiducial template power
spectrum assumes that a true acoustic scale as well as the shape of the
linear power spectrum is known a priori.

Here we investigate how the results depend on small deviations in the
template power spectrum.  We consider variations in $\Oh$, $\Obhh$,
and the spectral tilt $n$ of 2\%, chosen to be somewhat larger than
the expected uncertainties from Planck CMB anisotropy data.  When one
uses a different cosmological template compared to the true simulation
input, one of course measures a shift in $\alpha$.  However, this shift
in $\alpha$ is very close to that predicted by the ratio of the sound
horizons in the two cosmologies, where the sound horizon $r_s$ is given
by equation (6) in \citet{EH98}.  We find that $\alpha/r_s$ is recovered
to no worse than $\pm0.02$\% within the above cosmological variations,
as shown in Table \ref{tab:wrong}.

This implies that the measured acoustic scale is consistently determined,
independent of reasonable variations in the template cosmology.  Applying
this standard ruler to cosmological distance inferences, one will
consistently measure $D_A(z)/r_s$ and $H(z)r_s$ despite variations in the
template used.  Of course, when one tests a given cosmology against 
a suite of cosmological measurements, one must account for the change 
in $r_s$ in using the BAO constraints, just as one must self-consistently
alter any other predictions for cosmological observables.

\section{Conclusions}\label{sec:disc}
Baryon acoustic oscillations in large-scale structure can serve as
an excellent standard ruler to probe the acceleration history of the
Universe. Such a standard ruler method is based on the  premise that
the characteristic scale of the feature, the sound horizon scale, is
well measured from CMB and remains fixed throughout the evolution of
the Universe.

In this paper, we have measured the shift in the acoustic scale with high
signal-to-noise ratio using $320\trihGpc$ volume of the PM simulations. We
allow a large number of parameters in our fitting formula so that the
standard ruler method is dominated by the BAO.

We find nonlinear growth decreases the measured acoustic scale by less
than 1\%: we find shifts of 0.20\% at $z=1.5$, 0.26\% at $z=1$, 0.33\%
at $z=0.7$, and 0.45\% at $z=0.3$, all determined at 5 to 8$\sig$. In
detail, we find the redshift dependence of $\al-1$ scales as
$D^{1.66^{+0.26}_{-0.22}}$, where $D$ is the linear growth function.
The perturbation theory prediction of a $D^2$ scaling is consistent 
at 1.3 $\sig$.
We consider different fitting formula, polynomial orders, and resampling
methods and find highly consistent results between various choices.

Our detection of a sub-percent-level shift in the acoustic scale agrees with
the analytic node-by-node predictions of nonlinear mode-coupling shift
by \citet{Crocce08} within the same order of magnitude. The agreement is
better when we compare our results with the shifts they predict assuming
a global fit.

Redshift distortions increase the shift by $\sim 25\%$ relative to real
space. Using $216\trihGpc$ of simulations, we detect a shift of 0.25\%
at $z=1.5$, 0.33\% at $z=1$, 0.41\% at $z=0.7$, 0.54\% at $z=0.3$ at
the level of 5 to $6\sig$.

Although our results confirm the previous analytic prediction that
nonlinear growth and redshift distortions induce a small mode-coupling
shift on the acoustic scale, we find that the shift can be reduced to less
than 0.1\% by undoing the nonlinear growth using a simple reconstruction
scheme \citep{ESSS07}, even in the presence of non-negligible shot noise.

We compare our results from \Nb\ simulations with the analytic Fisher
matrix error forecasts based on \citet{SE07}. We find the standard
ruler method from the simulations and Fisher formalism forecasts are
in agreement within a modest level of sample variance. In detail, we
find that $\sial$ from the simulations is systematically better, by no
more than 27\%, than the Fisher matrix calculations.
While this trend could have an underlying physical cause,
it is consistent with sample variance, correlated between redshifts,
from our limited number of simulations.

We show that small variations in the template power spectrum around the
true fiducial cosmology do not affect our results of $\al/r_s$. Therefore,
the cosmological distance ratios $D_A(z)/r_s$ and $H(z)r_s$ inferred
from the measurements of the acoustic scale are robust to reasonable
errors in the template cosmology.

We argue that it is correct and appropriate to use the linear theory
power spectra as a template for extracting the acoustic scale.
The acoustic oscillations are not a perfect harmonic sequence nor
perfectly exponentially damped.  For example, because of Silk damping
\citep{Silk68},
small-scale perturbations in the baryons decouple from the CMB earlier
than large-scale perturbations, implying that the sound horizon is
not a single quantity.  However, the correct waveform is computed in
great detail in the Boltzmann codes.  Using an approximate waveform
will simply risk a loss of precision and accuracy.  The exact relation
of the acoustic signature and acoustic scale in the CMB to that in
the low-redshift large-scale structure depends upon the details of
recombination, but this is exquisitely constrained by CMB observations.
Once one has accepted the assumptions in interpreting the CMB anisotropies
to measure the densities needed to set the sound horizon, it is 
straight-forward to use the predicted waveform for the low-redshift power spectrum.

Finally, we emphasize that the nonlinear shifts on the acoustic scale
are predictable numerically, as shown here for each estimator. With such
a prediction, the shifts can be properly modeled and therefore can be
removed in the standard ruler analysis of actual clustering data, even
without reconstruction; the results can be crosschecked for consistency
with the results after reconstruction, when available.

We have successfully computed the effects of nonlinear growth and
redshift distortions on the acoustic scale using a large volume of
simulated Universe. The galaxy power spectrum is expected to exhibit
somewhat larger shifts than the matter power spectrum, as  galaxies would
sample more nonlinear regions. We will investigate this in a separate
paper (Siegel et al., in preparation). Encouraged by the successful
reduction of shift by a reconstruction technique applied to the mass
distribution with a nonzero shot noise, we will also investigate the
use of reconstruction of BAO in the galaxy power spectrum in future work.

\acknowledgements
We thank Martin Crocce for useful conversations. H.-J.S.\ is supported by
the DOE at Fermilab.
E.R.S., D.J.E., and M.W.\ were supported by NASA grant NNX07AH11G.
E.R.S.\ and D.J.E. were supported by NASA grant NNX07AC51G and 
NSF AST-0707725.
The simulations reported here used resources at
the National Energy Research Supercomputing Center.



\clearpage

\newcommand{\tableskip}{\\[-8pt]}
\newcommand{\singleline}{\tableskip\hline\tableskip}
\newcommand{\doubleline}{\tableskip\hline\tableskip}
\newlength{\tablespread}\setlength{\tablespread}{30pt}
\newcommand{\dje}{\hspace{\tablespread}}
\tabletypesize{\small}
\def\arraystretch{1.1}

\begin{deluxetable}{c@{\dje}c@{\dje}cccc}
\tablewidth{255pt}
\tablecaption{\label{tab:fullreal}The mean and the error of $\al$ in real space}
\startdata \doubleline
$z$&Model   &$\Signl$ &     $\al-1 (\%)$   &       $\sig_\al (\%)$  & $\rechi$  \\\singleline
0.3&Poly7       &7.6      &0.45        &0.055   &1.09   \\ 
0.3&Pade    &7.6     &0.45        &0.057    &1.00  \\
0.3&Poly3 &7.6        &0.43     &0.056          &1.24 \\\singleline

0.7&Poly7  &6.3 &0.33      &0.046      &1.10 \\
0.7&Pade  &6.3	    &0.33	    & 0.046   &0.97   \\ \singleline

1.0&Poly7  &5.5        &0.27         &0.041    &1.13  \\ 
1.0&Pade  &5.5	    &0.26	    &0.042    &0.99   \\ \singleline

1.5&Poly7 &4.5      &0.20           &0.037    &1.17  \\
1.5&Pade  &4.5	    &0.20	    &0.037    &1.04   \\\doubleline
0.3&Pade        &6.0    & 0.45      &0.057   &0.98 \\
0.3&Pade 	&7.0   	& 0.45      &0.057   &0.99   \\ 
0.3&Pade        &10.0   & 0.45      &0.057    &1.03 \tableskip
\enddata
\tablecomments{Fitting range: $0.02\ihMpc \leq k \leq 0.35\ihMpc$ (a total of 70 data points). Reduced $\chi^2$, $\rechi$, where ``DOF" is the number of degrees of freedom, is calculated from the best fit to the averaged power spectra over $320\trihGpc$. }
\end{deluxetable}

\begin{deluxetable}{c@{\dje}c@{\dje}cccc}
\tablewidth{310pt}
\tablecaption{\label{tab:fullred} The mean and the error of $\al$ in redshift space.}
\startdata \doubleline
$z$&Model   &$\Signl$ &     $\al-1 (\%)$   &    $\sial (\%)$      & $\rechi$ \\\singleline
0.3&Poly-Exp-Out     &9.0     &0.54        &0.070 (0.085)     &1.26  \\ 
&Poly-Exp-In      &9.0 &0.55  &0.070 (0.086) &1.25 \\
&Poly7                &9.0 &0.54  &0.068 (0.083) & 1.24\\
&Poly-Rat-In      &9.0 &0.50  &0.084 (0.103)  &1.46 \\
&Poly-Rat-Out       &9.0 &0.53  &0.069 (0.084) & 1.23\\
&Pade-Rat-In   &9.0 &0.55  &0.077 (0.093)&1.28 \\\singleline
0.7& Poly-Exp-Out      &8.0        &0.41	    &0.057 (0.069)   &1.12   \\ 
1.0& Poly-Exp-Out	&7.0	    &0.33	    &0.049 (0.060)   &1.10   \\ 
1.5&  Poly-Exp-Out     &6.0	    &0.25	    &0.041  (0.050)  &1.11      \\\doubleline
0.3&  Poly-Exp-Out     &8.0     &0.54        &0.071 (0.086)     &1.268 \\ 
0.3&  Poly-Exp-Out     &9.0     &0.54        &0.070 (0.085)     &1.256 \\ 
0.3&  Poly-Exp-Out     &10.0     &0.54        &0.070 (0.085)     &1.259 \\
0.3&  Poly-Exp-Out     &11.0     &0.57        &0.070 (0.085)     &1.413 \tableskip
\enddata
\tablecomments{Fitting range: $0.02\ihMpc \leq k \leq 0.35\ihMpc$. Values of $\sial$ from the \Nb\ (the fifth column)  are the errors for 40 simulations ($320\trihGpc$) rescaled from the errors for 27 simulations ($216\trihGpc$). The actual errors for 27 simulations are listed within the parentheses. Reduced $\chi^2$ is calculated from the best fit to the averaged power spectra over $216\trihGpc$.}
\end{deluxetable}

\clearpage
\begin{deluxetable}{c@{\dje}c@{\dje}cccc@{\dje}cccc}
\tablewidth{420pt}
\tablecaption{\label{tab:recreal} The effect of density-field reconstruction in real space.}
\startdata \doubleline
$z$ &Model& \multicolumn{4}{c@{\dje}}{Before Reconstruction} &\multicolumn{4}{c}{After Reconstruction} \\ \singleline
&      &$\Signl$ &     $\al-1 (\%)$   &       $\sig_\al (\%)$      & $\rechi$
&$\Signl$ &     $\al-1 (\%)$   &       $\sig_\al (\%)$      & $\rechi$ \\\singleline
0.3&Poly7   &7.6     &0.40        &0.071      &1.09                                &3.0     &0.07        &0.044      & 1.15\\
0.7&Poly7                       &6.3      &0.27       &0.062	 &1.07
    		&2.0     &0.03	&0.049	& 1.19 \\
1.0&Poly7         &5.5     &0.22         &0.075     &1.00
 &2.0     &0.00  &0.052&1.13 \\
1.5&Poly7         &4.5    &0.19          &0.079    &0.99
         	&2.0	 &0.03	&0.080	&1.04 \tableskip
\enddata
\tablecomments{Fitting range: $0.02\ihMpc \leq k \leq 0.5\ihMpc$ (a total of 481 data points). We use the 1\% sample. }
\end{deluxetable}

\begin{deluxetable}{c@{\dje}c@{\dje}cccc@{\dje}cccc}
\tablewidth{450pt}
\tablecaption{\label{tab:recred}The effect of density-field reconstruction in redshift space}
\startdata \doubleline
$z$ &Model& \multicolumn{4}{c@{\dje}}{Before Reconstruction} &\multicolumn{4}{c}{After Reconstruction} \\ \singleline
&      &$\Signl$ &     $\al-1 (\%)$   &       $\sig_\al (\%)$      & $\rechi$
&$\Signl$ &     $\al-1 (\%)$   &       $\sig_\al (\%)$      & $\rechi$ \\\singleline
0.3&Poly-Exp-Out      &9.0     &0.48         &0.079      &1.31 
                     &4.0     &0.07        &0.046      &1.20 \\ 
0.7&Poly-Exp-Out	   &8.0     &0.30        &0.081  &1.14  
                &4.0     &0.03	    &0.047 &1.21 \\
1.0&Poly-Exp-Out	  &7.0	    &0.31	    &0.093 &1.19 
 &3.0	    &0.01	    &0.056 &1.28 \\
1.5&Poly-Exp-Out          &6.0	    &0.27	    &0.079 &1.08  	   
&3.0	    &0.04	    &0.070 &1.19 \tableskip
\enddata
\tablecomments{Fitting range: $0.02 \ihMpc\leq k \leq 0.5\ihMpc$. We use the 1\% sample.  }
\end{deluxetable}

\clearpage
\begin{deluxetable}{l@{\dje}c@{\dje}ccc@{\dje}ccc}
\tablewidth{415pt}
\tablecaption{\label{tab:Fisher}Fisher Matrix Estimates in Comparison to Simulation Results}
\startdata \doubleline
& & \multicolumn{3}{c@{\dje}}{\Nb\ data} &\multicolumn{3}{c}{Fisher matrix} \\ 
\singleline
&$z$ &Sample &$\Signl$ &         $\sig_\al (\%)$    &$\Signn$  &  $\nPt$ & $\sial$ \\
\singleline
Real space & 0.3& 100\% &7.6           &0.055   
&7.6  &100 & 0.068\\

&0.7& 100\% &6.3            &0.046
&6.3  &100 & 0.054 \\

&1.0& 100\% &5.5            &0.041
&5.5 &100 & 0.047\\

&1.5& 100\% &4.5             &0.037
 &4.5 &100  &0.039\\
\singleline

&0.3& 1\%  &7.6              &0.071      
&7.6 &2.36 &0.084 \\

&0.7& 1\%  &6.3              &0.062
&6.3 &1.47 &0.076\\

&1.0& 1\%  &5.5              &0.075 
&5.5  &1.10 &0.075 \\

&1.5& 1\%  &4.5              &0.079 
&4.5  &0.72 &0.078 \\
\singleline

&$z$  &Sample & $\Signl$   &       $\sig_\al (\%)$      & $\Sigxy/\Sigz$ &  $\nPt(\mu=0)$ & $\sial$ \\ 
\singleline
Redshift space&0.3&100\%     &9.0         &0.070 (0.085)   
& 7.6/12.1 &100 & 0.085\\

&0.7&100\% &9.0	       &0.058  (0.071)
&6.3/11.0  &100 & 0.069\\

&1.0&100\% &7.0        &0.049  (0.060)
&5.5/10.1  &100 & 0.060\\

&1.5&100\% &6.0       &0.041  (0.050)
&4.5/8.6    &100  & 0.049\\ 
\singleline

&0.3 & 1\% & 9.0      &0.079 
& 7.6/12.1   &2.36 &0.10\\

&0.7&1\%     &8.0     &0.081  
&6.3/11.0   &1.47 &0.089\\

&1.0&1\%       &7.0    &0.093 
&5.5/10.1   &1.10 &0.084\\

&1.5&1\%         &6.0         &0.079
&4.5/8.6    &0.72 &0.081 \tableskip
\enddata
\tablecomments{Values of $\Signl$ in the fourth column represents the nonlinear smoothing used for the template P(k) in the $\chi^2$ analysis of the \Nb\ data, and $\Signn$ and $\Sigxy/\Sigz$ in the sixth column are derived from the \Zel\ approximations and represents the nonlinear degradation of the BAO assumed in the Fisher matrix calculations. For the 100\% samples in redshift space, we list both the actual $\sial$ from the 27 simulations (inside the parentheses) and the rescaled $\sial$ for 40 simulations.}
\end{deluxetable}

\begin{deluxetable}{lccc}
\tablewidth{280pt}
\tablecaption{\label{tab:wrong} The effect of a template $P_m$ with an incorrect cosmology}
\startdata \doubleline
& $z$ & Cosmology Alteration & Ratio of $\al/r_s$ \\ 
\singleline
Real space  & 0.3  & 1.02 $\Oh$ & 1.0002  \\
	    & 0.3  & 0.98 $\Oh$ & 0.9998  \\
	    & 0.3  & 1.02 $\Obhh$& 0.9999  \\
	    & 0.3  &$n_s=0.95$ & 1.0000 \\
	    & 1.5  &1.02 $\Oh$ & 1.0002\\
\singleline
Redshift space\hspace*{24pt}  & 0.3  & 1.02 $\Oh$ & 1.0001 \\
		& 1.5  & 1.02 $\Oh$ & 1.0002 \tableskip
\enddata
\tablecomments{The ratio of $\al/r_s$, where $r_s$ is the sound horizon,
in the fourth column compares the values of $\al/r_s$ from an incorrect
cosmology template with the one from the fiducial cosmology. Our
fiducial cosmology is $\Oh=0.1225$, $\Obhh=0.0224$, and $n=0.97$. Fitting
range: $0.02\ihMpc \leq k \leq 0.35\ihMpc$. We use the 100\% sample.}
\end{deluxetable}


\begin{thebibliography}{}\frenchspacing

\bibitem[Amendola et al.(2005)]{Amen05} Amendola, L.,
Quercellini, C., \& Giallongo, E.\ 2005, \mnras, 357, 429

\bibitem[Angulo et al.(2005)]{Angulo05} Angulo, R., Baugh,
C.~M., Frenk, C.~S., Bower, R.~G., Jenkins, A., \& Morris, S.~L.\ 2005,
\mnras, 362, L25 

\bibitem[Angulo et al.(2008)]{Angulo08} Angulo, R.~E., Baugh, 
C.~M., Frenk, C.~S., \& Lacey, C.~G.\ 2008, \mnras, 383, 755 

\bibitem[Bennett et al.(2003)]{BennettWmap} Bennett, C.~L., et al.\
2003, \apjs, 148, 1 


\bibitem[Beno{\^ i}t et al.(2003)]{BenoitArcheops} Beno{\^ i}t, A.~et
al.\ 2003, \aap, 399, L19

\bibitem[de Bernardis et al.(2000)]{deB00}
     de Bernardis, P., et al., 2000, Nature, 404, 955

\bibitem[Bharadwaj(1996a)]{Bhara96a} Bharadwaj, S.\ 1996, \apj, 
472, 1 

\bibitem[Bharadwaj(1996b)]{Bhara96b} Bharadwaj, S.\ 1996, \apj, 
460, 28 

\bibitem[Blake \& Glazebrook(2003)]{Blake03} Blake, C., \&
Glazebrook, K.\ 2003, \apj, 594, 665

\bibitem[Blake \& Bridle(2005)]{Blake05} Blake, C., \& Bridle,
S.\ 2005, \mnras, 363, 1329

\bibitem[Blake et al.(2006)]{Blake06} Blake, C., Parkinson, D.,
Bassett, B., Glazebrook, K., Kunz, M., \& Nichol, R.~C.\ 2006, \mnras, 365,
255

\bibitem[Blake et al.(2007)]{Blake07} Blake, C., Collister, A., 
Bridle, S., \& Lahav, O.\ 2007, \mnras, 374, 1527 

\bibitem[Bond \& Efstathiou(1984)]{Bond84}
        Bond, J.R. \& Efstathiou, G. 1984,
        \apj, 285, L45

\bibitem[Cole et al.(2005)]{Cole05} Cole, S., et al.\ 2005,
\mnras, 362, 505

\bibitem[Cooray(2004)]{Cooray04} Cooray, A.\ 2004, \mnras, 348,
250


\bibitem[Crocce \& Scoccimarro(2006)]{Crocce06b} Crocce, M., \&
Scoccimarro, R.\ 2006, \prd, 73, 063520

\bibitem[Crocce \& Scoccimarro(2008)]{Crocce08} Crocce, M., \& Scoccimarro, R.\ 2008, \prd, 77, 023533 


\bibitem[de Lapparent et al.(1986)]{deLa86} de Lapparent, V.,
Geller, M.~J., \& Huchra, J.~P.\ 1986, \apjl, 302, L1

\bibitem[Dolney et al.(2006)]{Dolney06} Dolney, D., Jain, B., \&
Takada, M.\ 2006, \mnras, 366, 884


\bibitem[Eisenstein \& Hu(1998)]{EH98} Eisenstein, D.~J., \&
Hu, W.\ 1998, \apj, 496, 605

\bibitem[Eisenstein et al.(1998)]{EHT98} Eisenstein, D.~J., 
Hu, W., \& Tegmark, M.\ 1998, \apjl, 504, L57 

\bibitem[Eisenstein(2003)]{Eisen03}
        Eisenstein, D.J., 2003, in ASP Conference Series, volume 280, Next Generation Wide Field Multi-Object Spectroscopy,
        ed. M.J.I. Brown \& A. Dey (ASP: San Francisco) pp. 35-43;
        astro-ph/0301623


\bibitem[Eisenstein et al.(2005)]{Eisen05} Eisenstein, D.~J.,
et al.\ 2005, \apj, 633, 560 

\bibitem[Eisenstein et al.(2007)]{ESSS07} Eisenstein, D.~J., 
Seo, H.-J., Sirko, E., \& Spergel, D.~N.\ 2007, \apj, 664, 675 

\bibitem[Eisenstein et al.(2007)]{ESW07} Eisenstein, D.~J., 
Seo, H.-J., \& White, M.\ 2007, \apj, 664, 660 

\bibitem[Estrada et al.(2008)]{Estra08} Estrada, J., Sefusatti, 
E., \& Frieman, J.~A.\ 2008, ArXiv e-prints, 801, arXiv:0801.3485 

\bibitem[Glazebrook \& Blake(2005)]{Glazebrook05} Glazebrook, K., \&
Blake, C.\ 2005, \apj, 631, 1

\bibitem[Guzik et al.(2007)]{Guzik07} Guzik, J., Bernstein, G., 
\& Smith, R.~E.\ 2007, \mnras, 375, 1329 

\bibitem[Halverson et al.(2002)]{HalDasi}
        Halverson, N.~W.~et al.\ 2002, \apj, 568, 38


\bibitem[Hamilton(1998)]{Hamilton98} Hamilton, A.J.S., 1998,
        ``The Evolving Universe'', ed. D. Hamilton (Kluwer: Dordrecht), p. 185;
        astro-ph/9708102  %


\bibitem[Hanany et al.(2000)]{Han00}
        Hanany, S., et al., 2000,
        \apj, 545, L5

\bibitem[Hinshaw et al.(2007)]{Hinshaw07} Hinshaw, G., et al.\ 
2007, \apjs, 170, 288 


\bibitem[Hinshaw et al.(2008)]{Hinshaw08} Hinshaw, G., et al.\ 
2008, ArXiv e-prints, 803, arXiv:0803.0732 




\bibitem[Holtzman(1989)]{Holtzman89} Holtzman, J.~A.\ 1989, \apjs,
71, 1


\bibitem[Hu \& Sugiyama(1996)]{HS96} Hu, W., \& Sugiyama,
N.\ 1996, \apj, 471, 542

\bibitem[Hu \& Haiman(2003)]{Hu03} Hu, W., \& Haiman, Z.\
2003, \prd, 68, 063004

\bibitem[Hu \& White(1996)]{Hu96} Hu, W., \& White, M.\ 1996, \apj, 471, 30 

\bibitem[Huff et al.(2007)]{Huff07} Huff, E., Schulz, A.~E., 
White, M., Schlegel, D.~J., 
\& Warren, M.~S.\ 2007, Astroparticle Physics, 26, 351 

\bibitem[H{\"u}tsi(2006)]{Hutsi06} H{\"u}tsi, G.\ 2006, \aap,
449, 891

\bibitem[Jain \& Bertschinger(1994)]{Jain94} Jain, B., \& Bertschinger, E.\ 1994, \apj, 431, 495 

\bibitem[Jeong \& Komatsu(2006)]{Jeong06} Jeong, D., \&
Komatsu, E.\ 2006, \apj, 651, 619

\bibitem[Juszkiewicz(1981)]{Jusz81} Juszkiewicz, R.\ 1981, 
\mnras, 197, 931 

\bibitem[Kaiser(1987)]{Kaiser87} Kaiser, N.\ 1987, \mnras, 227,
1 

\bibitem[Lee et al.(2001)]{Lee01} Lee, A.~T., et al.\ 2001,
\apjl, 561, L1

\bibitem[Linder(2003)]{Linder03} Linder, E.~V.\ 2003, \prd, 68,
083504 

\bibitem[Ma(2007)]{Ma07} Ma, Z.\ 2007, \apj, 665, 887

\bibitem[Makino et al.(1992)]{Makino92} Makino, N., Sasaki, M., 
\& Suto, Y.\ 1992, \prd, 46, 585 

\bibitem[Matarrese \& Pietroni(2007)]{Matarrese07} Matarrese, S., \& Pietroni, M.\ 2007, ArXiv Astrophysics e-prints, arXiv:astro-ph/0703563


\bibitem[Matsubara(2004)]{Mat04} Matsubara, T.\ 2004, \apj,
615, 573 

\bibitem[Matsubara(2008)]{Mat08} Matsubara, T.\ 2008, \prd, 
77, 063530 

\bibitem[Meiksin, White, \& Peacock(1999)]{Meiksin99}
        Meiksin, A., White, M., \& Peacock, J.~A.\ 1999, \mnras, 304, 851


\bibitem[Meiksin \& White(1999)]{MW99} Meiksin, A., \&
White, M.\ 1999, \mnras, 308, 1179

\bibitem[Miller et al.(1999)]{Mil99}
        Miller, A.D., Caldwell, R., Devlin, M.J., Dorwart, W.B., Herbig, T.,
        Nolta, M.R., Page, L.A., Puchalla, J., Torbet, E., \& Tran, H.T., 1999,
        \apj, 524, L1


\bibitem[Netterfield et al.(2002)]{Netter02} Netterfield, C.~B.,
et al.\ 2002, \apj, 571, 604

\bibitem[Nishimichi et al.(2007)]{Nishimichi07} Nishimichi, T., et
al.\ 2007, \pasj, 59, 1049

\bibitem[Okumura et al.(2008)]{Okumura08} Okumura, T., Matsubara, 
T., Eisenstein, D.~J., Kayo, I., Hikage, C., Szalay, A.~S., 
\& Schneider, D.~P.\ 2008, \apj, 676, 889 

\bibitem[Padmanabhan et al.(2007)]{Pad07} Padmanabhan, N., et 
al.\ 2007, \mnras, 378, 852 

\bibitem[Padmanabhan \& White(2008)]{Pad08} Padmanabhan, N., \& White, M.\ 2008, ArXiv e-prints, 804, arXiv:0804.0799 

\bibitem[Papai \& Szapudi(2008)]{Papai08} Papai, P., \& Szapudi, I.\ 2008, ArXiv e-prints, 802, arXiv:0802.2940 



\bibitem[Pearson et al.(2003)]{Pearson03} Pearson, T.~J., et al.\
2003, \apj, 591, 556

\bibitem[Peebles \& Yu(1970)]{Peebles70} Peebles, P.~J.~E.~\& Yu,
J.~T.\ 1970, \apj, 162, 815


\bibitem[Percival et al.(2007a)]{Percival07a} Percival, W.~J., et 
al.\ 2007, \apj, 657, 51 

\bibitem[Percival et al.(2007b)]{Percival07b} Percival, W.~J., Cole, 
S., Eisenstein, D.~J., Nichol, R.~C., Peacock, J.~A., Pope, A.~C., 
\& Szalay, A.~S.\ 2007, \mnras, 381, 1053 

\bibitem[Perlmutter et al.(1999)]{Perlm99} Perlmutter, S.~et
al.\ 1999, \apj, 517, 565
 
\bibitem[Riess et al.(1998)]{Riess98} Riess, A.~G.~et al.\
1998, \aj, 116, 1009

\bibitem[Sanchez et al.(2008)]{Sanchez08} Sanchez, A.~G., Baugh, C.~M., \& Angulo, R.\ 2008, MNRAS, submitted (arXiv:0804.0233)  

\bibitem[Scoccimarro et al.(1999)]{Sco99} Scoccimarro, R., 
Zaldarriaga, M., \& Hui, L.\ 1999, \apj, 527, 1 

\bibitem[Scoccimarro(2004)]{Sco04} Scoccimarro, R.\ 2004,
\prd, 70, 083007

\bibitem[Seljak et al.(2003)]{SSWZ} Seljak, U., Sugiyama,
N., White, M., \& Zaldarriaga, M.\ 2003, \prd, 68, 083507

\bibitem[Seljak \& Zaldarriaga(1996)]{SeZa96}
    Seljak, U., \& Zaldarriaga, M., 1996, \apj, 469, 437

\bibitem[Seo \& Eisenstein(2003)]{SE03} Seo, H.-J., \&
Eisenstein, D.~J.\ 2003, \apj, 598, 720

\bibitem[Seo \& Eisenstein(2005)]{SE05} Seo, H.-J., \&
Eisenstein, D.~J.\ 2005, \apj, 633, 575

\bibitem[Seo \& Eisenstein(2007)]{SE07} Seo, H.-J., \&
Eisenstein, D.~J.\ 2007, \apj, 665, 14


\bibitem[Silk(1968)]{Silk68} Silk, J.\ 1968, \apj, 151, 459


\bibitem[Smith et al.(2007)]{Smith07} Smith, R.~E.,
Scoccimarro, R., \& Sheth, R.~K.\ 2007, \prd, 75, 063512

\bibitem[Smith et al.(2008)]{Smith08} Smith, R.~E.,
Scoccimarro, R., \& Sheth, R.~K.\ 2008, \prd, 77, 043525


\bibitem[Springel et al.(2005)]{Springel05} Springel, V., et al.\
2005, \nat, 435, 629

\bibitem[Sunyaev \& Zeldovich(1970)]{SZ70} Sunyaev, R.~A., \& Zeldovich, Y.~B.\ 1970, \apss, 7, 3 

\bibitem[Takahashi et al.(2008)]{Taka08} Takahashi, R., et
al.\ 2008, ArXiv e-prints, 802, arXiv:0802.1808

\bibitem[Tegmark et al.(2006)]{Tegmark06} Tegmark, M., et al.\
2006, \prd, 74, 123507


\bibitem[Vishniac(1983)]{Vish83} Vishniac, E.~T.\ 1983, 
\mnras, 203, 345 

\bibitem[Wagner et al.(2007)]{Wagner07} Wagner, C., M{\"u}ller,
V., \& Steinmetz, M.\ 2007, arXiv:0705.0354 

\bibitem[White \& Scott (1996)]{Whi96} White, M., \& Scott, D. 1996, \apj, 459, 415

\bibitem[White(2005)]{White05} White, M.\ 2005, Astroparticle
Physics, 24, 334


\bibitem[Zhan \& Knox(2006)]{Zhan06} Zhan, H., \& Knox, L.\
2006, \apj, 644, 663

\bibitem[Zel'dovich(1970)]{Zel70}
        Zel'dovich, Y.A., 1970, \aap, 5, 84

\end{thebibliography}
\end{document}